\documentclass[useAMS,usenatbib]{mn2e}
\usepackage{epsfig}
\usepackage{amssymb}
\usepackage{amsmath}
\usepackage{lscape}
\usepackage{longtable}
\usepackage[normal]{caption}
\usepackage{appendix}
\usepackage{bm}

\def\arcsec{$^{\prime\prime}$}

\title[Linking the X-ray and infrared properties of star-forming
galaxies at $z<$1.5]{Linking the X-ray and infrared properties of
  star-forming galaxies at $z<$1.5\thanks{{\it Herschel} is an ESA space observatory with science instruments provided by European-led Principal Investigator consortia and with important participation from NASA.}}

\author[M.~Symeonidis et al.] 
{\parbox{\textwidth}{\raggedright
M.~Symeonidis,$^{1,2}$\thanks{E-mail: \texttt{msy@mssl.ucl.ac.uk}}
A.~Georgakakis,$^{3}$
M. J. ~Page,$^{2}$ 
J.~Bock,$^{4,5}$
M.~Bonzini,$^{6}$
V.~Buat,$^{15}$
D.~Farrah,$^{7}$
A.~Franceschini,$^{8}$
E.~Ibar,$^{19}$
D.~Lutz,$^{3}$
B.~Magnelli,$^{9}$
G.~Magdis,$^{10}$
S.J.~Oliver,$^{1}$
M. ~Pannella,$^{17,18}$
M. ~Paolillo,$^{11,16}$
D.~Rosario,$^{3}$
I.G.~Roseboom,$^{12}$
M.~Vaccari,$^{13}$
and C. ~Villforth$^{14,20}$ 
}\vspace{0.4cm}\\
\parbox{\textwidth}{\raggedright $^{1}$Astronomy Centre, Dept. of Physics \& Astronomy, University of Sussex, Brighton BN1 9QH, UK\\
$^{2}$Mullard Space Science
  Laboratory, University College London, Holmbury St. Mary, Dorking,
  Surrey RH5 6NT, UK\\
$^{3}$Max-Planck-Institut f\"ur Extraterrestrische Physik (MPE),Postfach 1312, 85741, Garching, Germany\\
$^{4}$California Institute of Technology, 1200 E. California Blvd., Pasadena, CA 91125, USA\\
$^{5}$Jet Propulsion Laboratory, 4800 Oak Grove Drive, Pasadena, CA
91109, USA\\
$^{6}$European Southern Observatory, Karl-Schwarzschild Str. 2, 85748
Garching bei Muenchen, Germany\\
$^{7}$Department of Physics, Virginia Tech, Blacksburg, VA 24061,USA\\
$^{7}$Laboratoire d'Astrophysique de Marseille, OAMP, Universit\'e
Aix-marseille, CNRS, 38 rue Fr\'ed\'eric Joliot-Curie, 13388 Marseille
cedex 13, France\\
$^{8}$Dipartimento di Astronomia, Universit\`{a} di Padova, vicolo
Osservatorio, 3, 35122 Padova, Italy\\
$^{9}$Argelander Institut fÃ¼r Astronomy, Bonn University, Auf dem HÃ¼gel 71, D-53121 Bonn, German\\
$^{10}$Department of Astrophysics, Denys Wilkinson Building,
University of Oxford, Keble Road, Oxford OX1 3RH, UK\\
$^{11}$Dipartimento di Fisica,  Universita di Napoli Federico II, C.U. di
Monte Sant'Angelo,  Via Cintia ed. G, 80126 Naples, Italy\\
$^{12}$Institute for Astronomy, Blackford Hill, Edinburgh EH9 3HJ, UK\\
Moffett Field, CA 94035\\
$^{13}$Astrophysics Group, Physics Department, University of the Western Cape, Private Bag X17, 7535, Bellville, Cape Town, South Africa\\
$^{14}$ School of Physics $\&$ Astronomy, Physical Science Building,
North Haugh, St Andrews, KY16 9SS, United Kingdom\\
$^{15}$Laboratoire d'Astrophysique de Marseille, OAMP, Universit\'e
Aix-marseille, CNRS, 38 rue Fr\'ed\'eric Joliot-Curie, 13388 Marseille
cedex 13, France\\
$^{16}$Agenzia Spaziale Italiana Science Data Center, Via del
Politecnico snc, 00133, Roma, Italy\\
$^{17}$Laboratoire AIM, CEA/DSM-CNRS-Université Paris Diderot, Irfu/Service Astrophysique, CEA-Saclay, 
Orme des Merisiers, 91191 Gif-sur-Yvette Cedex, France\\
$^{18}$Institut d'Astrophysique de Paris, UMR 7095, CNRS, UPMC Univ. Paris 06, 
98bis boulevard Arago, 75014, Paris, France\\
$^{19}$Instituto de F\'isica y Astronom\'ia, Universidad de
Valpara\'iso, Avda. Gran Breta\~na 1111, Valpara\'iso, Chile\\
$^{20}$SUPA, School of Physics and Astronomy, University of St
Andrews, North Haugh, St Andrews, KY16 9SS, UK
-\\
-\\
-\\
-\\
-\\
-\\
-\\
-\\
-\\
-\\
-\\
-\\
-\\
-\\
-\\
-\\
-\\
-\\
-\\
-\\
-\\
-\\
-\\
-\\
-\\
-\\
-\\
-\\
-\\
-
}}

\begin{document}

\date{Accepted  Received; in original form}

\pagerange{\pageref{firstpage}--\pageref{lastpage}} \pubyear{2014}

\maketitle

\label{firstpage}

\begin{abstract}
We present the most complete study to date of the X-ray emission from
star-formation in high redshift (median $z=0.7$; $z<$1.5), IR-luminous ($L_{\rm
  IR}=10^{10}-10^{13}$\,L$_{\odot}$) galaxies detected by
\textit{Herschel}'s PACS and SPIRE instruments. For our purpose we
take advantage of the deepest X-ray data to date, the Chandra deep
fields (North and South). Sources which host AGN are
removed from our analysis by means of multiple AGN indicators. We find an
AGN fraction of 18$\pm$2 per cent amongst our sample and note that AGN entirely dominate at values of log\,$[L_{\rm
  X}/L_{\rm IR}]>-3$ in both hard and soft X-ray bands. 
From the sources which are star-formation dominated,
only a small fraction are individually X-ray
detected and for the bulk of the sample we calculate average
X-ray luminosities through stacking. We find an average soft X-ray to
infrared ratio of log\,$\langle L_{\rm SX}/L_{\rm IR} \rangle=-4.3$
and an average hard X-ray to infrared ratio of log\,$\langle L_{\rm HX}/L_{\rm IR} \rangle=-3.8$.
We report that the X-ray/IR correlation is approximately linear through
the entire range of $L_{\rm IR}$ and $z$ probed and, although broadly consistent with
the local ($z<0.1$) one, it does display some discrepancies. We
suggest that these discrepancies are unlikely to be physical, i.e. due to an
intrinsic change in the X-ray properties of star-forming galaxies with
cosmic time, as there is no significant evidence for evolution of the
$L_{\rm X}$/$L_{\rm IR}$ ratio with redshift. Instead they are
possibly due to selection effects and remaining AGN
contamination. We also examine whether dust obscuration in the galaxy
plays a role in attenuating X-rays from star-formation, by
investigating changes in the  $L_{\rm X}$/$L_{\rm
  IR}$ ratio as a function of the average dust temperature. We
conclude that X-rays do not suffer any measurable attenuation in the
host galaxy.
\end{abstract}


\section{Introduction}
\label{sec:introduction}

From their onset, X-ray surveys have provided the most complete census of the
population of luminous ($L_X>10^{42}$\,erg\,s$^{-1}$) active galactic nuclei (AGN; e.g. Comastri et
al. 1995\nocite{Comastri95}; 2011\nocite{Comastri11}; Page et
al. 1997\nocite{Page97}; Hasinger et
al. 2005\nocite{Hasinger05}; Brandt $\&$ Hasinger 2005\nocite{BH05};
Tozzi et al. 2006\nocite{Tozzi06}; Tueller et
al. 2010\nocite{Tueller10}). Nevertheless, the X-ray detected population also includes a non-negligible
fraction of star-forming galaxies (SFGs), whose X-ray emission is
comparable to that of low luminosity AGN ($10^{38} \lesssim L_{X}
\lesssim 10^{42}$\,erg\,s$^{-1}$; eg. Griffiths $\&$ Padovani 1990\nocite{GP90}; David, Jones $\&$ Forman 1992\nocite{DJF92};
Ranalli, Comastri $\&$ Setti 2003\nocite{RCS03}; Franceschini et al. 2003\nocite{Franceschini03b}; Rosa-Gonz{\'a}lez et
al. 2007\nocite{Rosa-Gonzalez07}).
Although even the most luminous starburst galaxies are at least 3 and up to 5 orders
of magnitude less luminous in the X-rays than they are at optical and
infrared wavelengths (e.g. Franceschini et
al. 2003; Georgantopoulos et
al. 2005\nocite{GGK05}; U et al. 2012\nocite{U12}), X-rays allow us to
probe high energy processes in the interstellar medium (ISM) associated with
stellar evolution which are not accessible at other wavelengths.  X-ray emission is linked to a galaxy's star-formation history, chemical
evolution and ISM conditions (e.g. Ghosh $\&$ White 2001\nocite{GW01};
Fabbiano et al. 2004\nocite{Fabbiano04}), although it is neither dominated by direct starlight (like the UV
and optical), nor reprocessed starlight (like the infrared). The soft,
lower energy X-rays ($\lesssim$2\,keV) are primarily thermal
emission from gas in the ISM heated to X-ray temperatures by stellar
winds and supernovae, making them a good overall tracer
of the first $\sim$\,30\,Myr of star-forming activity (e.g. Mas-Hesse
et al. 2008\nocite{MOC08}). On the other hand,
more than 60 per cent of the hard high energy ($>$2\,keV) component is
resolved into point sources and associated with X-ray binaries (e.g.
Griffiths et al. 2000\nocite{Griffiths00}) and ultraluminous X-ray
sources (ULXs; e.g. Soria et al. 2010\nocite{Soria10};
2012\nocite{Soria12}; Fabbiano 2005\nocite{Fabbiano05}). High mass X-ray binaries
(HMXBs) in which the companion star is massive and short-lived, are
direct tracers of recent star-formation, whereas low mass X-ray
binaries (LMXBs) have lifetimes of the order of 10\,Gyr and are hence
more appropriate tracers of stellar mass (e.g. Ptak et
al. 2001\nocite{Ptak01}; Grimm et al. 2002, 2003\nocite{GGS02}\nocite{GGS03}). Such energetic
processes cannot be
probed at other wavelengths, making X-ray studies crucial for our
understanding of stellar and galaxy evolution. 

In the local Universe, where both resolution and sensitivity are in
one's favour, star-forming galaxies are particularly well studied in
the X-rays (e.g. Fabbiano $\&$ Trinchieri 1984\nocite{FT84}; Fabbiano
1988\nocite{Fabbiano88}; 1989\nocite{Fabbiano89}; Fabbiano et
al. 1997\nocite{FSM97}; Strickland et al. 2004\nocite{Strickland04};
Grimes et al. 2005\nocite{Grimes05}), even down to individual X-ray
sources.
As a result, X-ray emission has been evaluated as a tracer of star-formation
against other indicators such as infrared and radio 
(e.g. eg. Griffiths $\&$ Padovani 1990\nocite{GP90}; David et al. 1992\nocite{DJF92};
Ranalli et al. 2003\nocite{RCS03}; Franceschini et al. 2003\nocite{Franceschini03b}; Rosa-Gonz{\'a}lez et
al. 2007; Vattakunnel et al. 2012\nocite{Vattakunnel12}), although less often against the UV and optical because of dust extinction which
plagues that part of the spectral energy distribution (SED). After the
first infrared astronomical satellite (\emph{IRAS}) all
sky survey (Soifer et al. 1987\nocite{SNH87}) and the discovery of the most
luminous star-forming galaxies in the nearby Universe
(infrared-luminous galaxies; Soifer et al. 1984a\nocite{Soifer84a};
Sanders $\&$ Mirabel 1996\nocite{SM96}), it was established that IR
emission is an excellent tracer of the total star-formation rate (SFR)
in these infrared-bright galaxies as it corresponds to reprocessed emission of UV and optical
starlight by interstellar dust (e.g. Kennicutt
1998\nocite{Kennicutt98}). Since then, there have been a plethora of
studies targetting the link between the X-ray and infrared emission in
such luminous starburst systems at low redshifts (e.g. Grimm
et al. 2002; 2003; Persic et
al. 2004\nocite{Persic04}). In particular, X-ray and infrared emission
from the host galaxy share the  advantage of
being largely impervious to attenuation from dust and gas in the line of
sight as well as being orientation independent, unlike optical and UV star-formation tracers. 

Studies have shown that the X-ray/IR
correlation for star-forming galaxies is linear at high SFRs
(e.g. Gilfanov et al. 2004\nocite{GGS04}),
where both the X-rays and infrared emission directly trace the total
SFR, with a minimum contribution from the older stellar
population. The picture is less clear at lower SFRs, i.e. for
normal star-forming galaxies, where both the X-ray and IR
emission are thought to have a significant contribution from the older
stellar population, linking emission at those wavelengths to the stellar mass (e.g. Grimm et al. 2003\nocite{GGS03}; Gilfanov
et al. 2004; Colbert et al. 2004\nocite{Colbert04}; Lehmer et
al. 2010\nocite{Lehmer10}; Bendo et
al. 2010\nocite{Bendo10}; Lo Faro et al. 2013\nocite{LoFaro13}). As a result, there is evidence of
non-linearity in the X-ray/IR relation at low SFRs. Although for the
hard X-ray/IR relation this can be often attributed to higher contribution from
LMXBs in the integrated hard-band X-ray luminosity, the origin of non-linearity in the soft X-ray/IR correlation is a topic
of contention. It might be due to various factors such as gas in the line
of sight attenuating the soft X-rays, changes in density and
hence emissivity of the X-ray radiating gas (e.g. Grimes et
al. 2005\nocite{Grimes05}) or a consequence of stellar age (e.g. Mas-Hesse et al. 2008\nocite{MOC08}). 
As it currently stands, the local $z<0.1$ X-ray/IR correlation appears
overall non-linear over 4 orders of magnitude in SFR
(0.1-1000\,M$_{\odot}$/yr; e.g. see Symeonidis et al. 2011\nocite{Symeonidis11b}, hereafter S11). 

The picture is even less clear at high redshift, where only the
deepest X-ray surveys ($>$1\,Ms) detect X-ray emission from star-forming
galaxies and only from the most luminous of those (e.g. Ranalli et
al. 2005\nocite{RCS05}; S11). A significant complication is also AGN
contamination, as at high redshift it is currently not possible to distinguish whether X-rays originate from a low luminosity
AGN or a starburst galaxy. Moreover, comparisons between
local well-studied samples and high redshift samples is not trivial:
(i) IR-luminous galaxies are relatively rare at low redshifts, whereas their number density is
higher at high redshift; (ii) for the same $L_{IR}$, high
redshift sources display different dust properties to their local
equivalents (e.g. Coppin et al. 2008\nocite{Coppin08}; Sajina et al. 2008\nocite{Sajina08}; Farrah et
al. 2008\nocite{Farrah08}; Symeonidis et al. 2009\nocite{Symeonidis09}; 2013); (iii) weakly star-forming galaxies
are easily detected at low redshift, but often below the detection threshold of
high redshift surveys; (iv) we do not know how much of the
X-ray luminosity in log\,[$L_{\rm IR}$/L$_{\odot}$]\,$\gtrsim$\,12.5
sources originates in the host galaxy, rather than an
AGN, as these sources are too rare at $z<0.1$ where we could
potentially resolve the two components. 

In S11, we performed the
first study of X-ray emission from star-formation at $\langle z\rangle
\sim 1$ with a far-IR selected sample of galaxies detected by \textit{Herschel} (Pilbratt et
al. 2010\nocite{Pilbratt10}) in GOODS-N. Combining the small number of detections and retrieving average X-ray luminosities through
stacking for the remaining sources, we were able to evaluate the
X-ray to IR luminosity ratio ($L_{\rm X}$/$L_{\rm IR}$) of high
redshift SFGs against studies of equivalent sources in the local
Universe. We found that for luminous
and ultraluminous infrared galaxies (LIRGs and ULIRGs; $L_{\rm
  IR}$\,$>$\,10$^{11}$\,L$_{\odot}$), $L_{\rm X}$/$L_{\rm IR}$ was
consistent with values characteristic of local ($z<0.1$) equivalent sources,
indicating no evident evolution with redshift. In addition, we found
that the X-ray/IR correlation for star-forming galaxies could be taken
as linear in the high $L_{\rm IR}$ regime ($L_{\rm
  IR}$\,$>$\,10$^{11}$\,L$_{\odot}$). 

In this paper, we aim to re-visit this topic and advance the work we
presented in S11, by using a much larger
sample of galaxies (520) over both Chandra deep fields, as well
as the deepest X-ray data in those fields (2\,Ms in CDFN and 4\,Ms in
CDFS). We aim to probe the whole range of IR-luminous galaxies $L_{\rm
  IR}$\,$>$\,10$^{10}$\,L$_{\odot}$ and hence a large range of SFRs from $\sim$2 to 2000\,M$_{\odot}$\,yr$^{-1}$ (assuming the
conversion from $L_{\rm IR}$ to SFR by Kennicutt 1998). As this
sample is large and consists of the most intensely star-forming sources probed by
\textit{Herschel}, we expect to detect a non-negligible fraction of
them in the X-rays but also to achieve very high signal to noise through X-ray stacking. Our
goals are two-fold: (i) we intend to constrain the slope of the X-ray/IR
correlation over a large range in SFR and (ii) we intend to link their X-ray
properties to their infrared properties in order to gain a better
understanding of the physical nature of these systems. 

The paper is laid out as follows: Section \ref{sec:data} outlines the
data and sample selection, including the identification of AGN. In section \ref{sec:results} we
present our results and analysis. Finally our summary and conclusions can
be found in Section \ref{sec:conclusions}. Throughout we adopt a concordance cosmology of H$_0$=70\,km\,s$^{-1}$Mpc$^{-1}$, $\Omega_{\rm M}$=1-$\Omega_{\rm \Lambda}$=0.3.

\section{Sample selection}
\label{sec:data}

\subsection{Infrared observations}
\label{sec:infrared_data}

This work is based on \textit{Herschel} observations of the Great
Observatories Origins Deep Survey (GOODS)-North and South (Giavalisco
et al. 2002\nocite{Giavalisco04}) by \textit{Herschel}/PACS (Poglitsch et al. 2010\nocite{Poglitsch10})
as part of the PACS Evolutionary Probe survey (PEP; Lutz et al. 2011\nocite{Lutz11}) and
by \textit{Herschel}/SPIRE (Griffin et al. 2010\nocite{Griffin10}) as part of the \textit{Herschel} multi-tiered
extragalactic survey (HerMES; Oliver et al. 2012\nocite{Oliver12}). 
Source extraction in the PACS (100 and 160\,$\mu$m) and SPIRE (250,
350 and 500\,$\mu$m) bands is performed on the IRAC-3.6\,$\mu$m
positions of the f$_{24}$$\ge$30\,$\mu$Jy GOODS (N and S) sources, as described in Magnelli et
al. (2009\nocite{Magnelli09}) and Roseboom et
al. (2010\nocite{Roseboom10}; 2012\nocite{Roseboom12}); for information on the
GOODS \textit{Spitzer}/MIPS 24\,$\mu$m dataset see Magnelli et al. (2009\nocite{Magnelli09}). This method of
source extraction on prior positions is widely used and
enables identifications of secure counterparts over the whole SED. In this case however, its significant
advantage lies in its ability to effectively deal with source
blending in the \textit{Herschel} bands, particularly for
SPIRE where the beam is large (18.1, 24.9 and 36.6 arcsec FWHM at 250, 350 and
500\,$\mu$m respectively; Nguyen et al. 2010). By using prior
information to identify galaxies in the \textit{Herschel} images, we are able to
extract `clean' photometry for each galaxy, even for those which appear blended in the PACS and SPIRE bands.
The 3\,$\sigma$ sensitivity limits of the PACS 100 and 160\,$\mu$m
catalogues respectively are 3 and 6\,mJy for GOODS-N and 1 and 2\,mJy for
GOODS-S. For both GOODS fields, a 3\,$\sigma$ detection in
SPIRE using prior positions and the cross-identification method of
Roseboom et al. (2010) is approximately 8, 11 and 13\,mJy at 250, 350
and 500\,$\mu$m. In the case of the PACS bands $\sigma$ is only the photometric
error, whereas for the SPIRE bands, $\sigma$ includes confusion
error  (see Nguyen et al. 2010\nocite{Nguyen10} for the SPIRE
confusion limits).

\subsection{X-ray observations}
\label{sec:xray_data}

The X-ray data for GOODS-N are from the 2\,Ms \textit{Chandra} Deep
Field North (CDFN) survey (Alexander et al. 2003\nocite{Alexander03b}), with on-axis
sensitivity limits of $\sim$7.1$\times$10$^{-17}$\,ergs\,cm$^{-2}$\,s$^{-1}$ in the full (0.5-8
keV) band, $\sim$2.5$\times$10$^{-17}$\,ergs\,cm$^{-2}$\,s$^{-1}$ in the soft
(0.5-2.0 keV) band and $\sim$1.4$\times$10$^{-16}$\,ergs\,cm$^{-2}$\,s$^{-1}$ in the hard (2-8
keV) band. 
The X-ray data for GOODS-S are from the 4\,Ms observations of the CDFS
presented in Xue at al. (2011)\nocite{Xue11}, with on-axis sensitivity
limits of $\sim$\,3.2$\times$10$^{-17}$, $\sim$9.1$\times$10$^{-18}$ and $\sim$5.5$\times$10$^{-17}$
erg\,cm$^{-2}$\,s$^{-1}$ for the full (0.5-8\,keV), soft (0.5-2\,keV) and hard (2-8\,keV)
bands respectively. 

For uniformity purposes we use the final data products made available by the Imperial College (IC)
team\footnote{http://astro.ic.ac.uk/research/data-products-chandra-surveys}
--- see Laird et al. (2009\nocite{Laird09}) for details on the
methodology for data reduction, source detection and photometry
estimates. The
IC catalogues consist of X-ray sources with a Poisson probabilty that
the source is the result of random fluctuation of the
background of $<4\times10^{-6}$ (equivalent to $>$\,4.5$\sigma$
detections in the case of a
normal distribution), detected independently in four energy bands,
full (0.5--7\,keV), soft (0.5--2\,keV), hard (2--7\,keV) and ultra-hard
(5-7\,keV). We obtain fluxes and rest-frame luminosities in the 0.5--10 (full), 0.5--2 (soft) and 2--10\,keV (hard) energy band intervals, adopting a photon index of $\rm
\Gamma=1.9$, appropriate for SFGs. Hereafter, the subscripts FX, SX and HX
refer to the full, soft and hard X-ray bands.

\subsection{Initial sample selection}
\label{sec:sample}

The GOODS samples used in this work are
taken from Symeonidis et al. (2013\nocite{Symeonidis13a}; hereafter S13). The selection comprises all 24\,$\mu$m sources
that have detections (at least 3$\sigma$) at [100 and
160\,$\mu$m] OR [160 and 250\,$\mu$m] (where `OR' is the operator
representing disjunction in Boolean logic; i.e. it returns `true' if either or both
conditions are satisfied), thus obtaining a sample composed of dusty,
infrared-bright galaxies, the infrared luminosities and dust
temperatures of which can be robustly
measured. 

The redshifts we use are a combination of spectroscopic and
photometric, assembled from various catalogues: Berta et
al. (2011\nocite{Berta11}) for GOODS-N and Cardamone et
al. (2010\nocite{Cardamone10}) and Santini et al. (2009\nocite{Santini09}) for GOODS-S. The optical positions of sources in these catalogues are cross-matched to
the 24\,$\mu$m positions within 1\arcsec. The excellent photometric coverage of these fields and high quality
photometric redshifts available, result in $>$90 per cent of the
sources in our sample having a usable redshift. 
There are a total of 849 sources fulfilling the aforementioned
selection criteria (this is our IR-selected parent sample), 242 from GOODS-N, 62 per
cent of which have spectroscopic redshifts, and 607 from GOODS-S, 60
per cent with spectroscopic redshifts --- note that there are about 2.5 times more sources in GOODS-S as the
\textit{Herschel}/PACS data are deeper. 
Total infrared luminosities (8--1000\,$\mu$m; $L_{\rm IR}$) and average dust
temperatures for the sample are calculated as described in S13. $L_{\rm IR}$ is converted to SFR using the
Kennicutt (1998) relation. Stellar masses are determined with FAST
(Kriek et al. 2009\nocite{Kriek09}) using Bruzual $\&$ Charlot (2003\nocite{BC03})
delayed exponentially declining star formation histories (SFHs, $\psi(t) \propto \frac{t}{\tau^2} exp(-t/\tau)$) with 0.01$<\tau<$10
Gyr, solar metallicities, Salpeter initial mass function, and the
Calzetti 2000\nocite{Calzetti00} reddening law with $A_{V}$ up to 4
mag) --- more details in Pannella et al. in prep. 

We cross-match the positions of X-ray sources in the CDFN and CDFS
IC catalogues to the 3.6\,$\mu$m IRAC positions of our \textit{Herschel} sample within
2\,arsec, finding an X-ray detection rate (detection in at least one
X-ray band) of 31 per cent in CDFN and 22 per cent
in CDFS; in total 25 per cent of our sample is X-ray detected at the
$>4.5\sigma$ level. Note that although the GOODS-S survey is deeper in
the X-rays, it is also deeper in the infrared, hence we do not
necessarily expect a higher X-ray detection rate in the former.

\begin{figure}
\epsfig{file=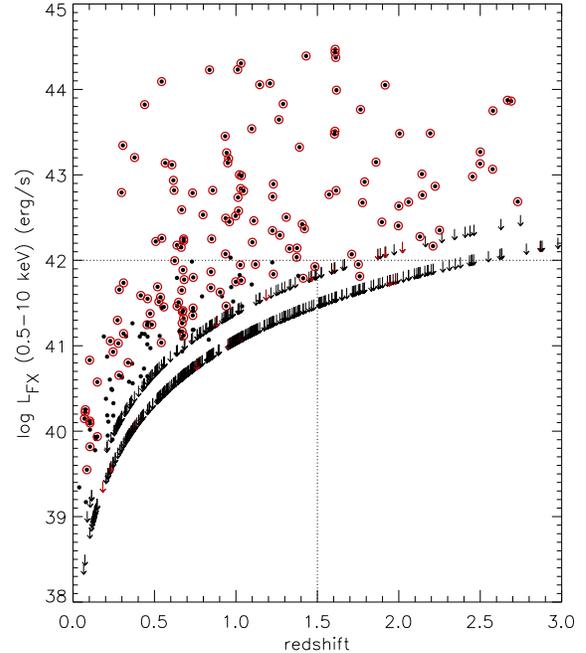, width=0.99\linewidth} 
\caption{The full band X-ray luminosity versus redshift for the
  IR-selected parent sample. AGN are indicated in red. The horizontal line marks 
  $L_{\rm SX}$=10$^{42}$\,erg\,s$^{-1}$: all sources with $L_{\rm SX}>$10$^{42}$\,erg\,s$^{-1}$
  are identified as AGN. The vertical line marks the redshift (z=1.5)
  above which these X-ray surveys are broadly insensitive to X-rays from
  star-formation. There are two sets of upper limits as the CDFS and
  CDFN surveys reach different depths. }
\label{fig:LX_z}
\end{figure}

\begin{figure}
\epsfig{file=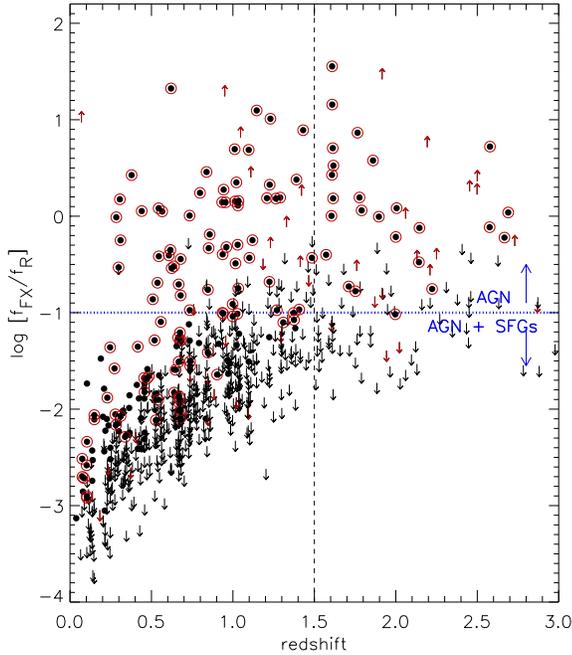, width=0.99\linewidth} 
\caption{The X-ray to R-band (log\,[$f_{\rm FX}$/$f_{\rm R}$]) ratio versus redshift for the
  IR-selected parent sample. AGN are indicated in red. The horizontal line marks the 
   log\,$[f_{\rm FX}$/$f_{\rm R}]=-1$ point above which all sources are
  classified as AGN, whereas at log\,$[f_{\rm FX}$/$f_{\rm R}]<-1$ both AGN
  and SFGs reside. The vertical line marks the redshift (z=1.5)
  above which only AGN can be identified with this criterion for the
  given flux limits.}
\label{fig:fxfopt}
\end{figure}

\subsection{Identification of AGN}
\label{sec:AGN}

To identify AGN in the IR-selected parent sample we use the following criteria:

\begin{itemize}

\item  $\bm f_{\rm \bm FX}$/$ \bm f_{\rm \bm R}$ {\bf{ratio}}: The full band X-ray to R-band flux ratio
  ($f_{\rm FX}$/$f_{\rm R}$) has been extensively used to separate
AGN and starburst systems since early observations of spectroscopically identified AGN have shown them to dominate the
-1$<$\,log\,[$f_{\rm FX}$/$f_{\rm R}$]$<$1 parameter space, with star-forming galaxies having values of log\,[$f_{\rm FX}$/$f_{\rm R}$]$<$-1 and typically $<$-2 (e.g. Hornschemeier et al. 2002\nocite{Hornschemeier02}, 2003\nocite{Hornschemeier03}; Akiyama et al. 2003\nocite{Akiyama03}; Georgantopoulos, Georgakakis $\&$ Koulouridis 2005\nocite{GGK05};
Georgakakis et al. 2007\nocite{Georgakakis07}). In this work, X-ray detected sources
with log\,$[f_{\rm FX}/f_{\rm R} ]>$-1 are assumed to be AGN hosts. 

\item {\bf{Hardness ratio}}: Hardness ratio (HR) is defined as $\frac{H-S}{H+S}$,
where $H$ and $S$ refer to the hard (2-7 keV) and soft (0.5-2\,keV)
count rates (counts\,s$^{-1}$). Although both star-forming
  systems and unabsorbed AGN are characterised by soft X-ray spectra
  and hence low hardness ratios (HRs), absorbed AGN have hard spectra and
  high values of hardness ratio, as low frequency X-rays are more
  severely attenuated by gas in the line of
  sight. Measured photon indices ($\Gamma$) of starburst galaxies
  range between 3 and 1.2 (e.g. Franceschini et al. 2003; Lehmer et
  al. 2010), corresponding to hardness ratios between -0.65 and -0.1
  and therefore sources with a hard band detection and HR$>$-0.1 are considered to host absorbed AGN.  

\item  {\bf{IRAC colours}}: \textit{Spitzer}/IRAC (3.6, 4.5, 5.8, 8\,$\mu$m) colours have been extensively
  used to identify AGN which are powerful enough to dominate the
  near/mid-IR emission their host. Their signature emerges as a power-law
  continuum in the near/mid-IR over what is normally an
  inflection in the SED of a star-forming galaxy. We use the
Donley et al. (2012\nocite{Donley12}) IRAC colour
($f_{5.8}$/$f_{3.6}$) - colour
($f_{8}$/$f_{4.5}$) criteria to identify AGN dominating the
near/mid-IR part of the SED --- see also S13.

\begin{figure}
\epsfig{file=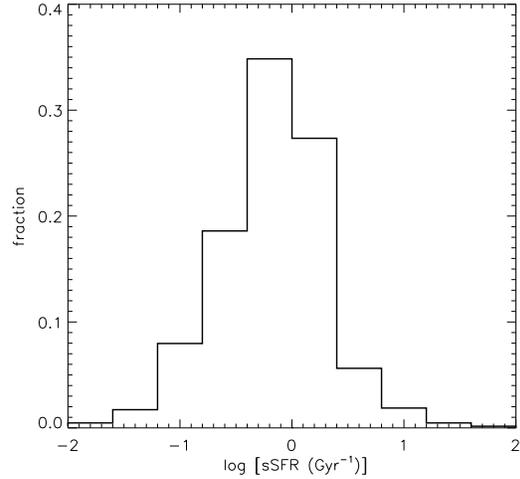, width=0.9\linewidth} 
\caption{The distribution in specific SFR (Gyr$^{-1}$) of our working sample
  of 640 sources at $z<1.5$ and with $L_{\rm IR}>$10$^{10}$\,L$_{\odot}$. The histogram is normalised to the total
  number of sources. }
\label{fig:sSFR}
\end{figure}

\item {\bf{Optical variability}}: 
Optical variability detected on timescales from hours to decades
originates from the nuclear region and is thus
used as an AGN criterion (e.g. Ulrich et al. 1997; Sarajedini et al. 2003). 
Optically variable AGN were identified using the catalogues of
Villforth et al. (2010\nocite{Villforth10} and
2012\nocite{Villforth12}) for the GOODS fields. Villforth et al. selected
variable sources through their flux in deep ACS
F850LP imaging and further analyzed the candidate AGN using multiwavelength data.

\item {\bf{X-ray variability}}: 
Although individual X-ray sources in star-forming galaxies are
variable, the integrated X-ray emission is not (e.g. Young et
al. 2012\nocite{Young12}) and thus for unresolved
galaxies X-ray variability is an AGN signature. We use the variability
catalogues of Paolillo et al. (2004\nocite{Paolillo04}; 2014 in prep.) and Young et al. (2012), to select X-ray variable
sources at the $>95\%$ probability level. 

\item {\bf{Spectroscopic identification}}: Narrow-line or broad-line
  AGN were isolated according to the spectral classifications
  in Szokoly et al. (2004),
  Mignoli et al. (2005), Treister et
  al. (2006), Ravikumar et al. (2007), Vanzella et al. (2005; 2006; 2008),
  Silverman et al. (2010) and 
  Balestra et al. (2010) for GOODS-S and Barger et al. (2005) and
  Treister et al. (2006) for GOODS-N. 

\item {\bf{Radio-loudness}}: For sources which are detected in
  the VLA 1.4GHz surveys of GOODS-N and GOODS-S, we identify the ones
  which are 3$\sigma$ above the radio-IR correlation shown in Seymour et
  al. (2011\nocite{Seymour11}), as hosting radio-loud AGN.

\end{itemize}

\begin{table}
\centering
\caption{Table showing the fraction of AGN recovered by each of the 8
  criteria, amongst the 216 X-ray detected sources (column 1) and
  amongst the IR-selected parent sample (column 2). 
The last row shows the total AGN fraction out of the X-ray detected
sources (column 1) and the IR-selected parent sample (column 2). }
\begin{tabular}{l|l|l|}
\hline 
& X-ray detected sources & Parent sample\\
\hline 
 $f_{\rm FX}$/$f_{\rm R} $ ratio &42$\%$ & 11$\%$\\
Hardness ratio& 34$\%$&9$\%$ \\
IRAC colours & 13$\%$& 3$\%$\\
Optical variability & 7$\%$& 2$\%$\\
X-ray variability & 29$\%$& 7$\%$\\
Radio-loudness & 7$\%$& 2$\%$\\
Optical spectra & 26$\%$& 7$\%$\\ 
\hline
Total & 82$\%$& 21$\%$\\
\hline
\end{tabular}
\label{table:AGN}
\end{table}

\begin{figure*}
\centering
\epsfig{file=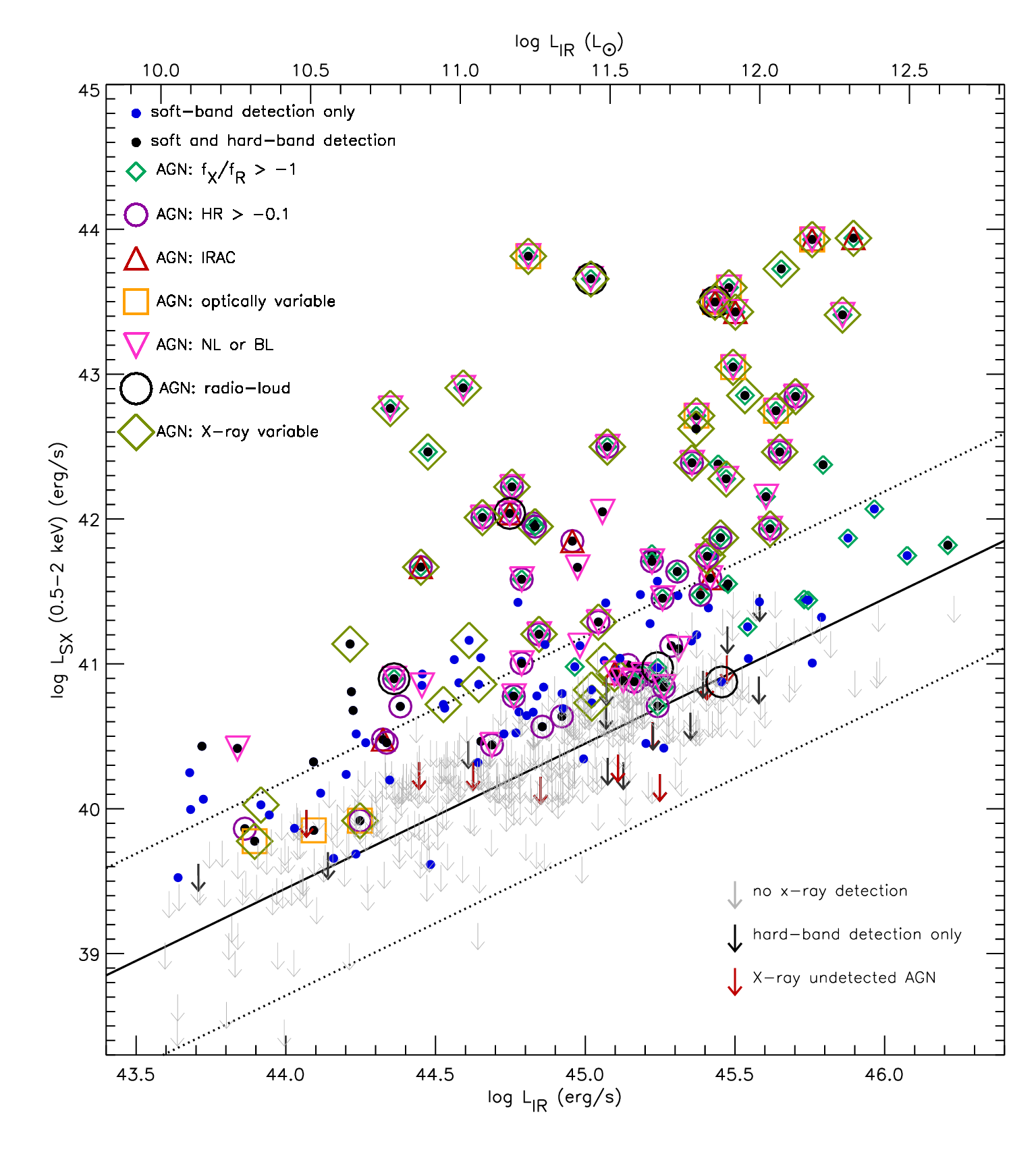, width=0.77\linewidth}
\caption{Plot of soft X-ray luminosity ($L_{\rm SX}$)  versus total
  infrared luminosity ($L_{\rm IR}$) for our working sample. As also shown in
  the legend: black filled circles denote sources which are detected in both
  soft and hard bands, whereas blue filled circles denote sources
  which are only soft-band detected. AGN are identified according to
  the criteria in section \ref{sec:AGN} and shown with different symbols
  here --- see legend. Grey upper limits are for sources with no
  X-ray detection and black upper limits are for sources with a
  hard-band detection. Red upper limits are X-ray undetected AGN hosts
  (AGN identified by other means). The black
     solid and dashed lines are the local \textit{soft} X-ray-IR
      star-formation correlation for $L_{\rm IR}>10^{11}$ sources from
      S11 (log\,$L_{\rm X}$=log\,$L_{\rm
        IR}$-4.55) and $\pm$2$\sigma$ boundaries ($\pm$0.74\,dex),
      here extrapolated to
      $L_{\rm IR}=10^{10}$. }
\label{fig:IR_Lxray_soft}
\end{figure*}

\begin{figure*}
\centering
\epsfig{file=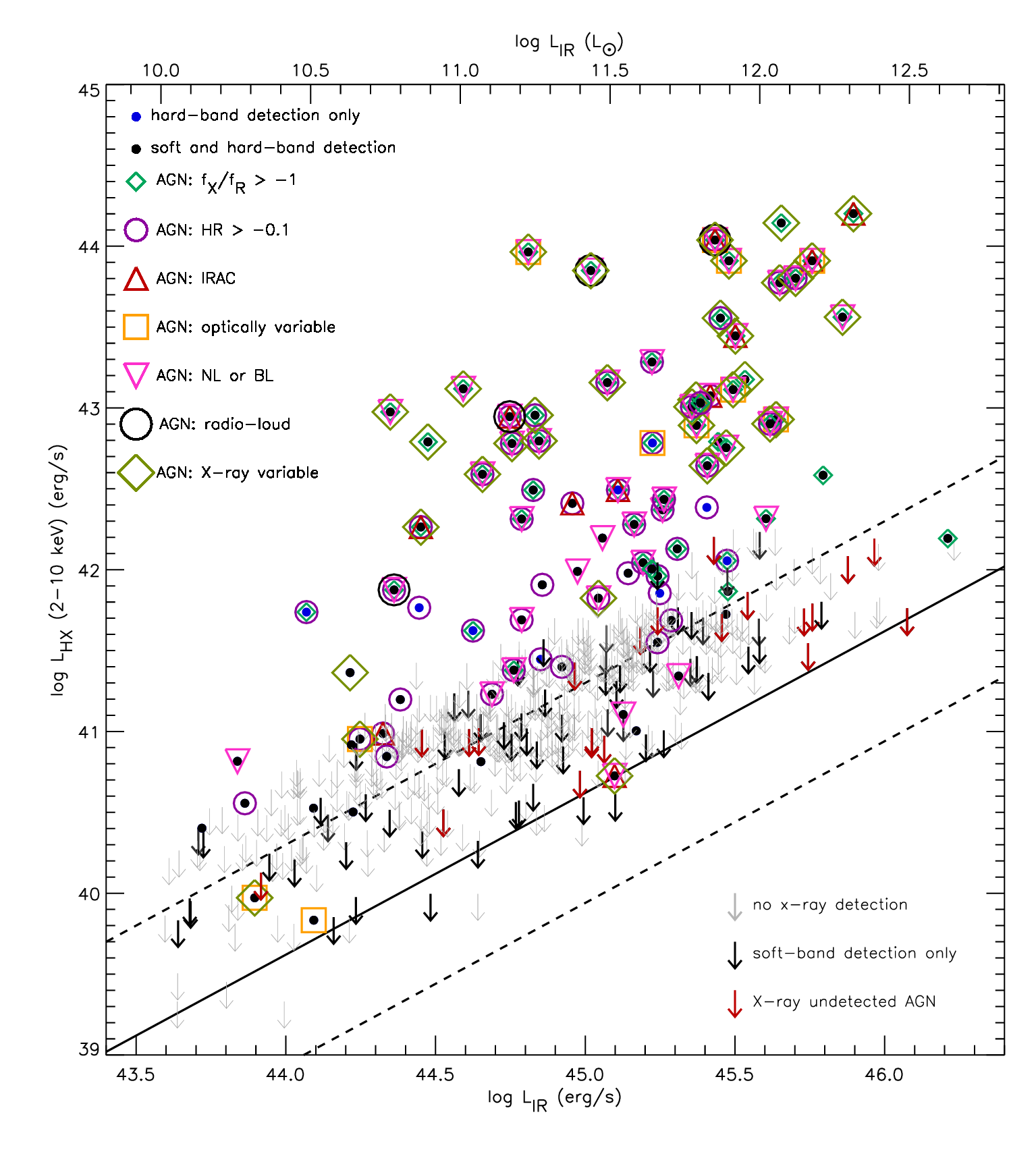, width=0.77\linewidth}
\caption{Plot of hard X-ray luminosity ($L_{\rm HX}$)  versus total
  infrared luminosity ($L_{\rm IR}$) for our working sample. As also shown in
  the legend: black filled circles denote sources which are detected in both
  soft and hard bands, whereas blue filled circles denote sources
  which are only hard-band detected. AGN are identified according to
  the criteria in section \ref{sec:AGN} and shown with different symbols
  here --- see legend. Grey upper limits are for sources with no
  X-ray detection and black upper limits are for sources with a
  soft-band detection. Red upper limits are X-ray undetected AGN
  hosts (AGN identified by other means). The black
     solid and dashed lines are the local \textit{hard} X-ray-IR
      star-formation correlation for $L_{\rm IR}>10^{11}$ sources from
      S11 (log\,$L_{\rm X}$=log\,$L_{\rm
        IR}$-4.38) and $\pm$2$\sigma$ boundaries
      ($\pm$0.68\,dex), here extrapolated to
      $L_{\rm IR}=10^{10}$. }
\label{fig:IR_Lxray_hard}
\end{figure*}

A total of 177 AGN are identified. Table \ref{table:AGN} shows the AGN recovered by each criterion
as a fraction of the 216 X-ray detected sources and out of the IR-selected parent sample. Most AGN are recovered through their X-ray
to optical flux ratio, hardness ratio, X-ray variability and
optical spectra, whereas only a small
fraction display radio-loudness, optical variability and IRAC
colours typical of AGN. This is not surprising as only a small
fraction ($\sim$10) per cent of AGN are commonly found to be radio-loud
and only the most powerful AGN in our sample of dust-rich galaxies will emerge in the IRAC colour-colour
diagram.

\subsection{Final sample}
\label{sec:final_sample}

In order to select the final sample used in this work we need to
minimise contamination from AGN. The obvious start is to remove all
identified AGN. However, since most galaxies are not
detected in the X-rays, our work will rely heavily on X-ray stacking
(see section \ref{sec:stacking} for details) and thus we also aim to
minimise contamination from luminous AGN in the X-ray undetected sources which
will form the bulk of our final sample. 
Fig. \ref{fig:LX_z} shows the $L_{\rm FX}-z$ parameter space of our
IR-selected parent sample. We see that (i) above $L_{\rm FX}$=10$^{42}$\,erg\,s$^{-1}$, all X-ray detected sources are
classified as AGN with our criteria and (ii) at $z>1.5$, the surveys'
limits are around $L_{\rm FX} \sim 10^{42}$\,erg\,s$^{-1}$, suggesting
that at $z>1.5$, $\le$4\,Ms X-ray surveys are
broadly insensitive to X-ray emission from star-formation. A similar
conclusion emerges from Fig. \ref{fig:fxfopt} which shows the log\,[$f_{\rm FX}$/$f_{\rm R}$] ratio as a function of
redshift. Above z$\sim$1.5, this criterion has identified AGN but no
SFGs, as the limiting X-ray flux of these surveys scatters around log\,$[f_{\rm FX}$/$f_{\rm R}]
\sim$-1, which is the value used to isolate the high luminosity AGN (see section \ref{sec:AGN}).   
Figs \ref{fig:LX_z} and \ref{fig:fxfopt} indicate that many of the AGN that can be identified
at low redshift, would be just below the detection threshold
at high redshift and as a result it is possible that they will
contaminate or even dominate the stacking signal. In addition, the
$f_{\rm FX}$/$f_{\rm R}$ criterion which is most effective in separating
AGN and SFGs (see table \ref{table:AGN}) at low redshift, does not constitute a fair
test above $z\sim$1.5, as due to the surveys' sensitivity most
sources have log\,$[f_{\rm FX}$/$f_{\rm R}] > -1$. 
In order to mitigate these issues, we exclude sources at $z>$1.5 from our final sample.  

The final cut is related to $L_{\rm IR}$; we remove sources with $L_{\rm IR}<$10$^{10}$ $L_{\odot}$ in
order to focus on IR-luminous galaxies, for which the bolometric
energy output peaks in the infrared. This ensures high star-formation rates and
thus less contamination from LMXBs in the hard
X-rays. Fig. \ref{fig:sSFR} shows the distribution in specific SFR (sSFR)
of our sample.
According to Mineo et al. (2012a\nocite{MGS12a}) above an sSFR of
$\sim$10$^{-10}$\,yr$^{-1}$ ($\equiv$ log\,[sSFR (Gyr$^{-1}$)]=-1), HMXBs
which are tracers of star-formation,
are expected to dominate the hard X-ray emission. Lehmer et al. (2010)
report a similar value for the sSFR of 5.6$\times$10$^{-11}$\,yr$^{-1}$ ($\equiv$
log\,[sSFR (Gyr$^{-1}$)]=-1.25). In this context, the sSFR distribution
for our sample (Fig. \ref{fig:sSFR}) shows that the vast majority of
sources can be considered HMXB-dominated
in the hard X-rays. 

Our working sample now consists of 640 sources at $z<1.5$ and with $L_{\rm IR}>$10$^{10}$\,L$_{\odot}$; Figs \ref{fig:IR_Lxray_soft} and
\ref{fig:IR_Lxray_hard}  show the soft and hard X-ray luminosity as a
function of total infrared luminosity. We note the following: most AGN are identified by more than one
criterion, (ii) there are some AGN which are not X-ray detected, 
(iii) the majority of X-ray detected sources are classified as
hosting AGN and (iv) most X-ray detected SFGs, are only detected in
the soft band, as the hard band flux limit is much shallower for both
CDF surveys. 

Within the errors, the fraction of AGN in our working sample (18$\pm$2 per
cent, where the error is binomial at 68 per cent) is consistent with the
fraction of 27$\pm$10 per cent we obtained in S11, where we examined
the AGN fraction of a small sample of IR-luminous galaxies in GOODS-N. Note that here, better statistics
have allowed us to decrease the uncertainty and hence more accurately constrain
the AGN fraction. However, as our AGN selection criteria are not complete, it is expected that some AGN have been
missed and hence this fraction is a lower limit. 

Our final sample, on which most of the ensuing analysis is based,
consists of the 524 star-forming galaxies (SFGs) which have not been identified as AGN
hosts by any criterion.

\subsection{Stacking in the X-rays}
\label{sec:stacking}
For the sources not individually detected in the X-rays, we use
stacking analysis to retrieve mean X-ray observed fluxes; below we briefly outline
our stacking methodology, but refer the reader
to Georgakakis et al. (2008\nocite{Georgakakis08a}) for more details. Stacking is performed in bins which contain 10 or more sources; see section
\ref{sec:results} and Figs \ref{fig:LIR_z} and \ref{fig:LIR_temp} for
the binning of the sample.  
We use an aperture radius of 2\arcsec to extract X-ray
photons at the IRAC 3.6\,$\mu$m positions of our sample. To account
for the remaining flux outside the 2 arcsec radius, we compute a mean
aperture correction by averaging the exposure-time weighted PSF corrections for
individual sources. X-ray sources in the IC CDF X-ray catalogues (see
section \ref{sec:xray_data}), were excluded, as were sources
that were close to an X-ray detection, i.e. by less than 1.5 times the local 90 per cent
encircled energy fraction (EEF) radius, where photons associated with
the wings of the PSF of the X-ray detections would contaminate the
signal. Moreover, as mentioned earlier, sources which are not X-ray detected but
identified as AGN are also excluded. 
The significance of the stacked signal is calculated relative
to the background within a 50 arcsec radius, subsequently scaled to
the area within the extraction aperture. For the background estimation
and to avoid contamination
we masked regions around X-ray detections using a radius
1.5 times larger than the 90 per cent EEF. Subsequently, in all figures we show the stacked luminosities as stars if
they are $>3\sigma$ and as upper limits otherwise. 

To verify that luminous but X-ray undetected AGN are not significantly altering the
signal in the stacking, we repeated the stacking after first removing
all detections down to 3$\sigma$ (as opposed to 4.5$\sigma$ which
is the depth of our X-ray catalogue, see
Section \ref{sec:xray_data}), i.e. performed stacking on the $<3\sigma$
sources. We found that the stacking signal did not change
significantly, as might have been the case if there were many AGN
just below the 4.5$\sigma$ threshold. As a result, for our analysis we
revert to the stacking as described above where only $\gtrsim
4.5\sigma$ sources are removed as this leaves a larger number of
sources per bin and hence better signal to noise.

\section{Results and Analysis}
\label{sec:results}

\begin{figure}
\begin{tabular}{c}
\epsfig{file=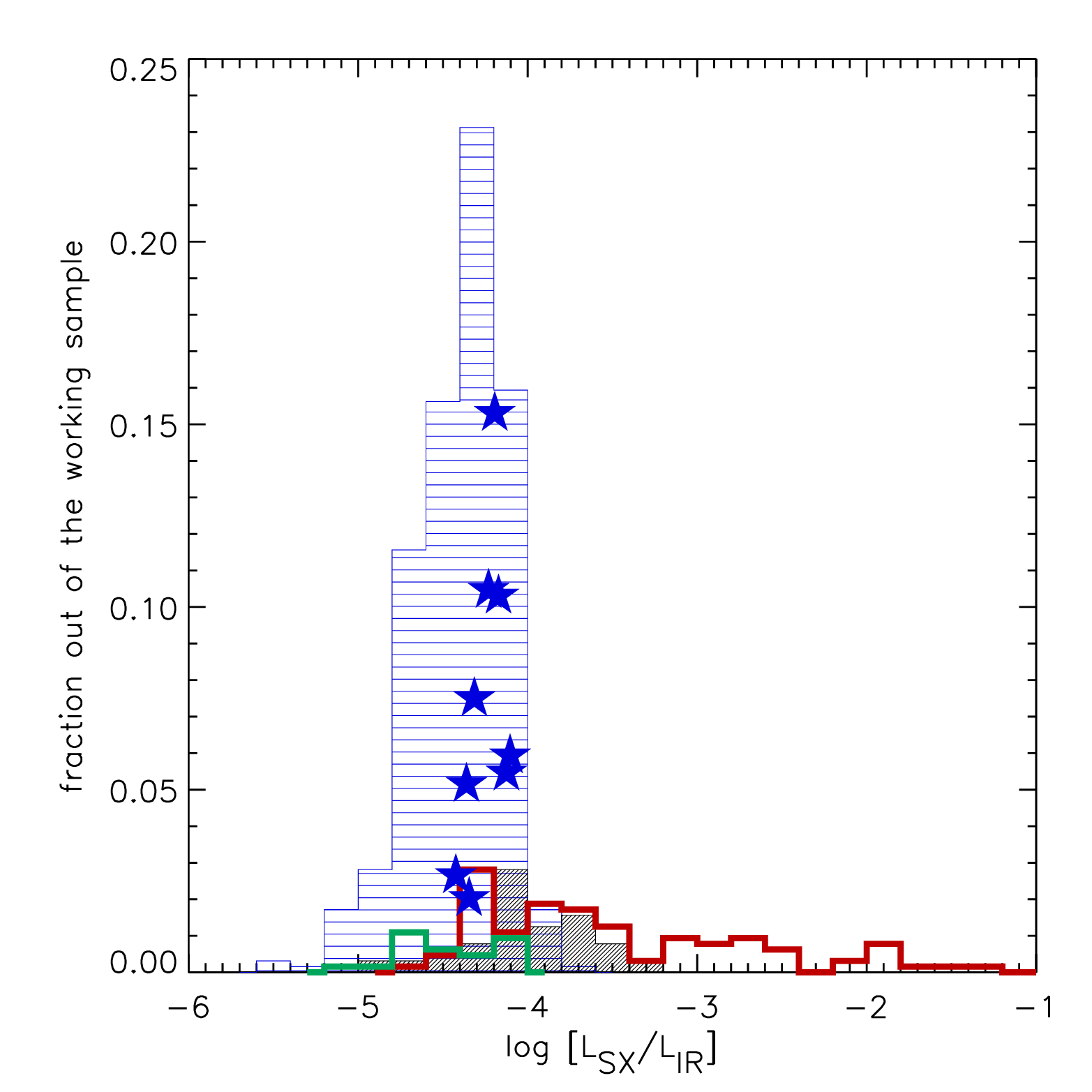, width=0.94\linewidth} \\
\epsfig{file=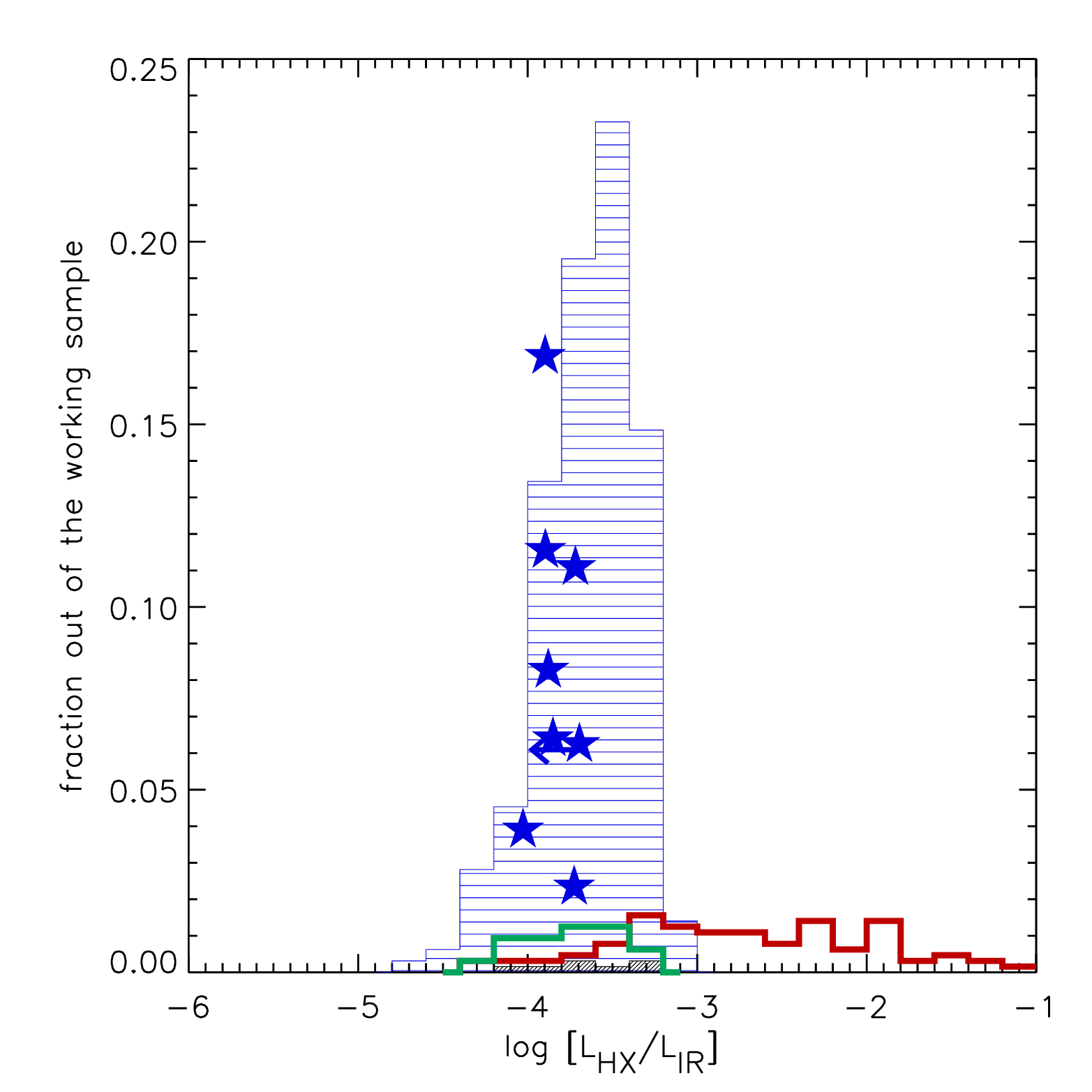, width=0.94\linewidth} \\
\end{tabular}
\caption{ The distribution in $L_{\rm SX}/L_{\rm
  IR}$ (top panel) and $L_{\rm HX}/L_{\rm
  IR}$ (lower panel) for our working sample of \textit{Herschel} objects, including the AGN. The lined blue
histogram is the distribution of sources which are not detected in the
X-rays, but which are assumed to have an X-ray luminosity equal to
their upper limits. The stacked values (see Fig. \ref{fig:LIR_z}
 for binning) are shown as blue stars if they
are $>3\sigma$ and as blue upper limits otherwise. The grey hatched histogram
are the X-ray detected SFGs (not classified as AGN) --- note that this
is barely visible in the lower panel. Finally, the red
histogram are the X-ray detected AGN, whereas the green
histogram is the X-ray undetected AGN assumed to have an X-ray
luminosity equal to their upper limit. Note that only AGN exist at
log\,$[L_{\rm X}/L_{\rm IR}]>$-3 in both X-ray bands.} 
\label{fig:ratio_histogram}
\end{figure}

\subsection{The distribution in $L_{\rm X}/L_{\rm
  IR}$ }
Fig. \ref{fig:ratio_histogram} shows the distribution in $L_{\rm X}/L_{\rm
  IR}$ for our working sample of \textit{Herschel} objects, including the AGN (for the X-ray
undetected sources we use the flux limit to convert to a luminosity). Interestingly, the
average luminosities from stacking are much closer to the detection limit in the soft band than in
the hard band, suggesting that we would only need a small increase in
sensitivity in the soft band in order to detect these sources. 
The AGN (detected and undetected) extend over the entire available range in $L_{\rm X}/L_{\rm
  IR}$, whereas the X-ray sources not identified as AGN hosts,
span a smaller range dropping to zero above log\,$L_{\rm X}/L_{\rm
  IR}$=-3. At low $L_{\rm X}/L_{\rm IR}$, there is significant overlap
in the distributions of AGN hosts and X-ray detected sources not
identified as AGN, however at
log\,$L_{\rm X}/L_{\rm IR}>$-3 (in both bands), the balance tips to
favour AGN making it possible to select `clean', AGN-dominated
samples in the X-rays, setting the X-ray/IR ratio as an additional
AGN/SFG criterion to select luminous AGN. 
On the other hand, it is not possible to disentagle the
AGN and star-formation components in sources hosting low luminosity AGN, particularly in samples such as ours where
galaxies are a priori known to be strongly star-forming. As a result,
it is likely that some sources not classified as AGN hosts
by our criteria, do host AGN, although the balance between the
star-formation and AGN emission in these sources is unclear. 

Including the stacking and detections for the SFGs, we measure  the
log of the (weighted) average \,$L_{\rm
  SX}/L_{\rm IR}$ ratio to be -4.3 and the log of the (weighted) average
\,$L_{\rm HX}/L_{\rm IR}$ ratio to be -3.8 (see also table \ref{table:SFGs}).
Using the Kennicutt (1998) relation to convert from $L_{\rm IR}$ to SFR,
this translates to 1.2$\times$10$^{39}$ erg\,s$^{-1}$ per unit SFR (M$_{\odot}$/yr) in
the soft band and 3.3$\times$10$^{39}$ erg\,s$^{-1}$ per unit SFR (M$_{\odot}$/yr) in the hard
band. Mineo et al. (2012a\nocite{MGS12a}; 2012b\nocite{MGS12b}) find for local, SFR$<$20\,M$_{\odot}$/yr
sources, $L_{\rm   SX} /\rm SFR \sim 8.3 \times$10$^{38}$ and $L_{\rm HX}
/\rm SFR \sim 2.6\times$ 10$^{39}$ erg/s per unit SFR (M$_{\odot}$/yr). They compare their
measurements to other studies and report that there exists a scatter
of a factor of 2--5 in the values reported, likely of physical origin. Note that although our values are
consistent with those emerging from other studies, we are probing much
higher SFRs.

\begin{table}
\centering
\caption{Table summarising the properties of the SFGs in our sample;
  the median redshift, the median total IR luminosity, the average
  soft X-ray/IR ratio and the average hard X-ray/IR ratio.}
\begin{tabular}{c|c|c|c|}
\hline 
median $z$&log median $L_{\rm IR}$&log\,$\langle L_{\rm SX}/L_{\rm
  IR}\rangle$&log\,$\langle L_{\rm HX}/L_{\rm IR}\rangle$\\
0.73&11.2&-4.3&-3.8\\
\hline
\end{tabular}
\label{table:SFGs}
\end{table}

\begin{figure}
\epsfig{file=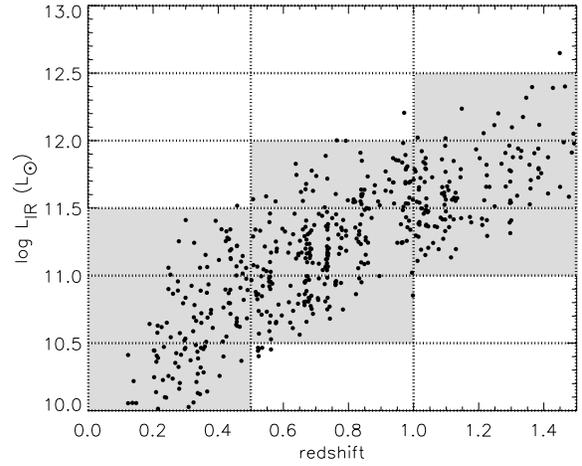, width=0.99\linewidth} 
\caption{Total infrared luminosity versus redshift for the
  \textit{Herschel} SFGs. X-ray stacking of
  undetected sources is performed in the
  indicated shaded $L_{\rm IR}-z$ bins which contain 10 or more undetected
  sources. X-ray detected SFGs are not included in the
  stacking. }
\label{fig:LIR_z}
\end{figure}

\begin{figure}
\begin{tabular}{c}
\epsfig{file=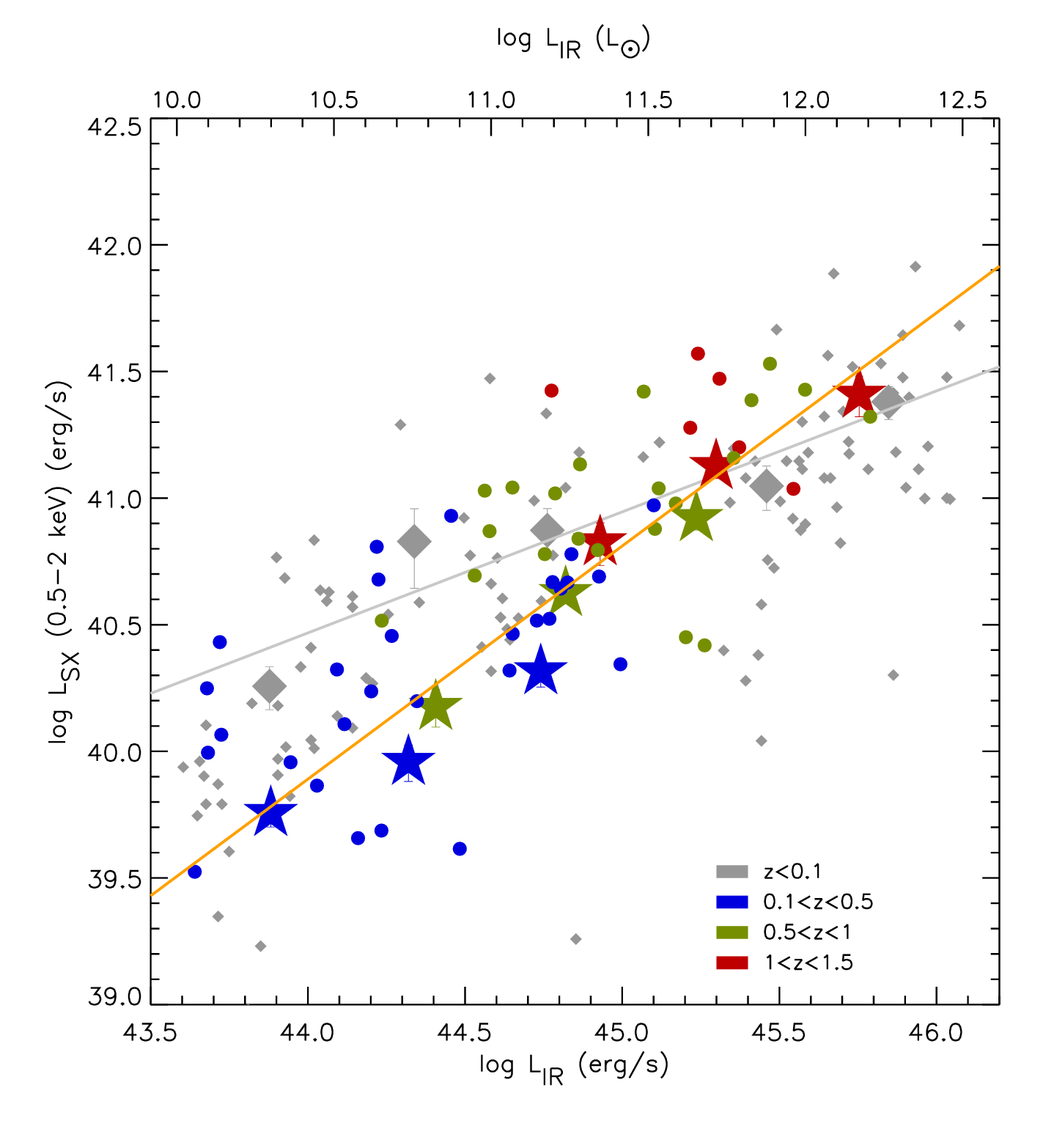, width=0.92\linewidth} \\
\epsfig{file=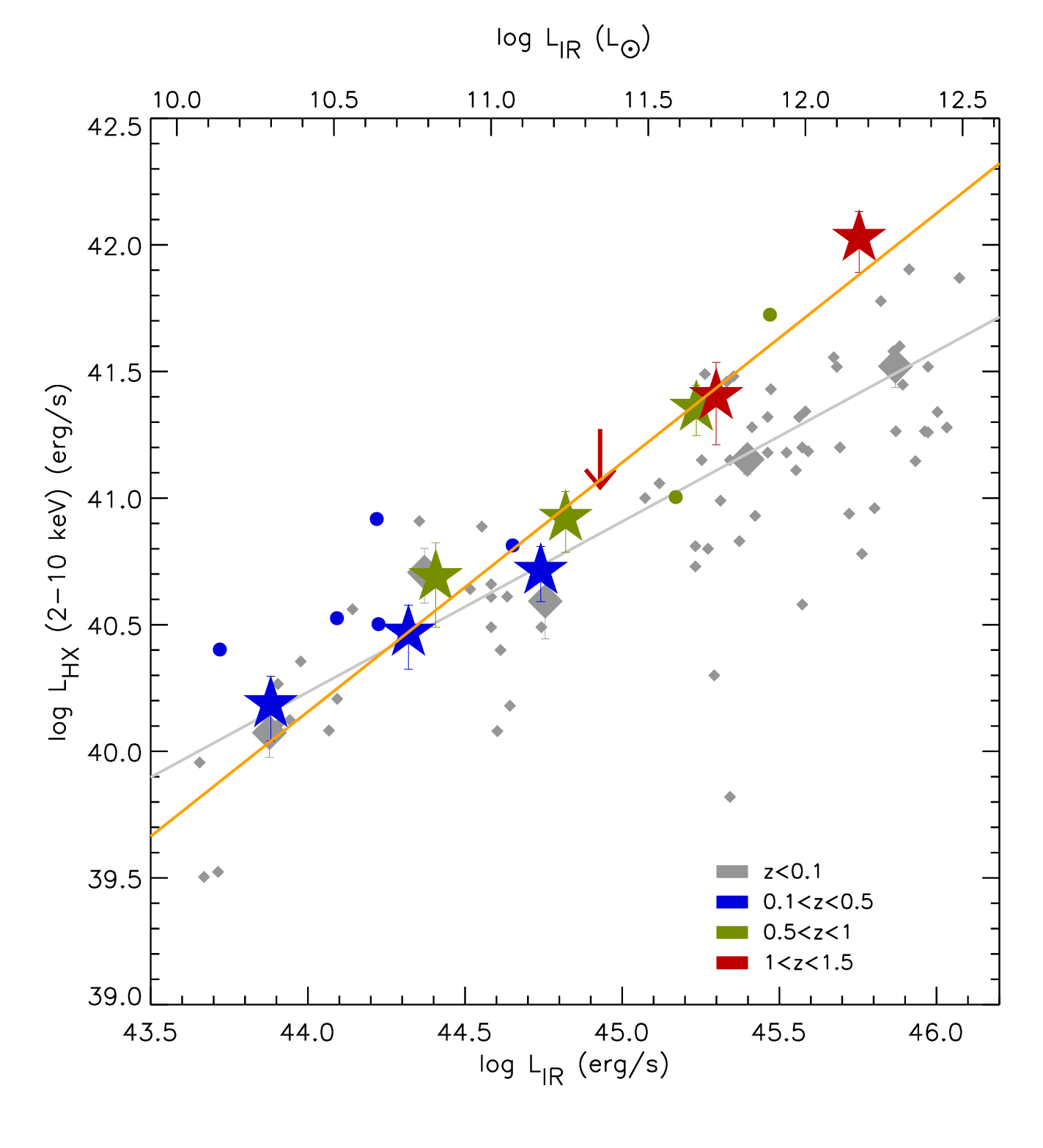, width=0.92\linewidth} \\
\end{tabular}
\caption{ Upper panel: soft X-ray luminosity ($L_{\rm SX}$) versus
  $L_{\rm IR}$. Lower panel: hard X-ray luminosity ($L_{\rm HX}$) versus
  $L_{\rm IR}$. In both panels, the  coloured symbols are our sample of \textit{Herschel} SFGs, blue for
$0.1<z<0.5$, green for 0.5$<z<$1 and red for 1$<z<$1.5. Large filled stars
correspond to the stacked values if they are $>3\sigma$ (otherwise
shown as upper limits), whereas
  the points correspond to the detections. The orange line is a log
  linear fit to the arithmetic weighted mean (in 5 $L_{\rm IR}$ bins) of the stacking and
  detections for the \textit{Herschel} SFGs. Note that because there
  are many more X-ray undetected sources per $L_{\rm IR}-z$ bin, the
  slope and normalisation of the best fit line are mainly determined by the
  stacks. For
  the \textit{Herschel} sample, the best fit equation of the soft
  X-ray/IR correlation (orange line in upper panel) is log\,$L_{\rm SX}$=0.93\,log$L_{\rm
    IR}$ - 0.82 and of the hard X-ray/IR correlation (orange line in lower panel) is log\,$L_{\rm HX}$=0.98\,log$L_{\rm
    IR}$ - 3.14. For both relations luminosities are in erg/s. 
The grey small diamonds are the local ($z<0.1$)
  sample taken from S11, with the grey line representing a log-linear
fit to the arithmetic mean (large grey diamonds) in 5 bins. The
equations of the local X-ray/IR correlations are
log\,$L_{\rm HX}$=0.48\,log$L_{\rm IR}$ + 19.4 in the soft band (grey line in upper panel)
and log\,$L_{\rm HX}$=0.67\,log$L_{\rm IR}$ + 10.6 in the hard band (grey line in lower panel). For both relations luminosities are in erg/s. } 
\label{fig:corr}
\end{figure}

\subsection{The X-ray/IR correlation for star-forming galaxies}
\label{sec:xray_ir_corr}

Fig. \ref{fig:LIR_z} shows the $L_{\rm IR}-z$ parameter space probed in this
work. The shaded bins are the ones used to examine the X-ray/IR
correlation (Fig. \ref{fig:corr}). The X-ray/IR correlation at high redshift was also
previously examined in S11, however only in the $L_{\rm IR} \sim
10^{11} - 10^{13}$\,L$_{\odot}$ range and with a sample roughly 10
times smaller than our current sample. In S11 we showed that the
X-ray/IR correlation for local galaxies is non-linear in the $L_{\rm IR} \sim
10^{10} - 10^{13}$\,L$_{\odot}$ range (see grey line in Fig. \ref{fig:corr}), but can be considered linear
when focusing on the higher luminosity sources $L_{\rm IR} >10^{11.5}$\,L$_{\odot}$ regime. We also found that the X-ray/IR ratio for $L_{\rm IR} >
10^{11}$\,L$_{\odot}$ sources at $z\sim$1 was consistent with that of
their local counterparts. 
Here, with a much larger sample, we can probe lower infrared luminosities ($L_{\rm IR} <
10^{11}$\,L$_{\odot}$) and SFRs, obtain
more reliable mean X-ray luminosities through stacking and revisit the
X-ray/IR correlation.

Fig. \ref{fig:corr} shows that the locus of most sources (individual detections and stacking) in the $L_{\rm SX} - L_{\rm IR}$ and
$L_{\rm HX} - L_{\rm IR}$ plane is within the range covered by the
local sample, however there are some discrepancies in the
average $L_{\rm X}$ per bin. Local ULIRGs have a
lower $L_{\rm X}/L_{\rm IR}$ ratio than normal IR galaxies (NIRGs,
10$^{10}$$<$$L_{\rm IR}$$<$10$^{11}$), whereas the high redshift sample does
not display the same behaviour. The \textit{Herschel} SFGs follow a quasi linear
$L_{\rm X} - L_{\rm IR}$ relation over 3 orders of
magnitude, offset from the local relation at low $L_{\rm
  IR}$ in the soft band and at high $L_{\rm
  IR}$ in the hard band. Indeed, when computing the weighted
arithmetic average of the stacking and detections in each $L_{\rm IR}$ bin and
fitting a log-linear relation to the average values, the 
 log\,$L_{\rm SX}$-log\,$L_{\rm IR}$ relation has a slope of
0.93$\pm$0.04 and the log\,$L_{\rm HX}$-log\,$L_{\rm IR}$
relation a slope of 0.98$\pm$ 0.09. In both cases, these numbers are
different from those obtained for the local sample, where the slope is calculated to be 0.48$\pm$0.1 in the soft band and
0.67$\pm$0.10 in the hard band.

In the local Universe, discrepancies in the hard
X-ray emission between galaxies of different SFRs is attributed to additional LMXB contribution boosting the
X-ray luminosity of low SFR sources (e.g. see extensive analysis in
Lehmer et al. 2010). However, some authors argue that obscuration
could also play a role in lowering the X-ray emission of the most
actively star-forming sources, where denser gas in the line
of sight might attenuate even the hard X-rays (e.g. see
Iwasawa et al. 2009; Lehmer et al. 2010). In the soft band, these differences could arise
because of large dispersion in stellar ages, as younger
systems are expected to show decreased X-ray luminosity per unit
SFR (e.g. Mas-Hesse et al. 2008\nocite{MOC08}). On the other hand,
changes in the distribution and emissivity of the hot X-ray gas could also be
responsible for the X-ray deficiency of local ULIRGs (e.g. Grimmes et
al. 2005). However, there is also the caveat that the local samples are not
selected homogeneously (the selection varies from optical to IR to
X-ray) and they are between 60 and 100 per cent complete. This makes them somewhat biased towards the 
brightest X-ray sources, particularly in the low SFR regime where
galaxies are less X-ray luminous, and could potentially serve to flatten the
slope of the local relation.

Now let us discuss the behaviour of the high redshift
\textit{Herschel} sample, with the local sample as our baseline
reference. At low $L_{\rm IR}$, we note a discrepancy in soft X-ray emission between the local and high redshift samples. If the reason behind these discrepancies is physical then it could be related to younger stellar ages or more extended soft X-ray emission in the high redshift sources; perhaps $z\sim0.1-0.5$ IR-luminous galaxies are already different to their
$z<0.1$ counterparts, at least with respect to their X-ray
properties. However, the reason could instead be due to a selection
bias, and we refer back to the caveat mentioned above. In contrast to
the local samples, our sample is complete, as the X-ray luminosity from all our sources is
accounted for and hence our analysis includes a large fraction of faint X-ray emitters perhaps
missing from some of the local studies. Another reason that
could account for these discrepancies is that, due to the redshift
range covered by the \textit{Herschel} sample, our X-ray data probes
harder rest-frame energies. Although for the local sample
the soft band mainly probes emission from hot gas, for the
\textit{Herschel} sample, even in the low redshift bin, we are likely
missing a large part of this hot gas emission and probing more of the
HMXB contribution. However, this does not explain why there is agreement in the
$L_{\rm SX}/L_{\rm IR}$ ratio for the high $L_{\rm IR}$ sources but not for
the low $L_{\rm IR}$ sources.

\begin{figure*}
\epsfig{file=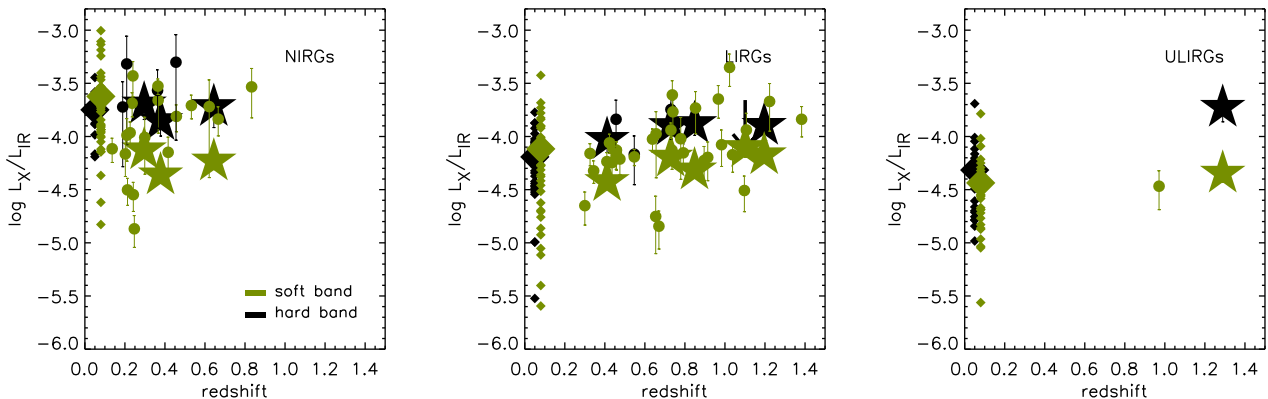, width=0.99\linewidth} 
\caption{The X-ray/IR luminosity ratio as a function of redshift for
  the \textit{Herschel} SFGs and local sample. For the
  \textit{Herschel} sample, large filled stars
correspond to the stacked values if they are $>3\sigma$ (otherwise
shown as upper limits) and filled circles are the X-ray
detections. For the local sample we plot small diamonds at an
arbitrary redshift of $<0.1$ with the large
  diamonds corresponding to the arithmetic mean. The colour-coding corresponds to
  the hard X-ray band (black) and the soft X-ray band (green). The 3 panels are
  the different $L_{\rm IR}$ luminosity classes, NIRGs, LIRGs and
  ULIRGs.} 
\label{fig:ratio}
\end{figure*}

In the hard band the differences between the two samples are mainly seen at
high $L_{\rm IR}$. Since in the hard band we probe the same
rest-frame energies for the local and high redshift samples, as the
simple power-law \textit{K}-correction performed is sufficient in this case, the
discrepancy is perhaps due to AGN contamination in the \textit{Herschel}
sample. In principle this would be more of a problem in the high redshift bin, where the survey detection
threshold selects more X-ray luminous sources (see Fig. \ref{fig:LX_z}). In the Appendix
we present a simple way to investigate the AGN impact and conclude that the AGN
contribution is minimal in the soft band, but more pronounced in the
hard band. In essence, we predict that $L_{\rm HX}$ for the $z>1$ bins
could be up to 0.26\,dex lower, under the assumption that AGN could be
boosting the average $L_{\rm HX}$. Although this would result in
better agreement between the high redshift and local relations, it
would also make the $L_{\rm HX} - L_{\rm IR}$ relation for the \textit{Herschel} sample
similarly non-linear to the local one. This is
surprising, because one would expect a linear relation if HMXBs trace star-formation, unless there is an additional
contribution from LMXBs (e.g. Lehmer et al. 2010; Mineo
et al. 2012a). Hence, it is possible that at low $L_{\rm IR}$, even
the \textit{Herschel} sources' X-ray luminosities are boosted
due to the contribution from LMXBs, despite their high sSFRs; perhaps the LMXB
contribution as a function of sSFR has been underestimated.

\begin{figure}
\epsfig{file=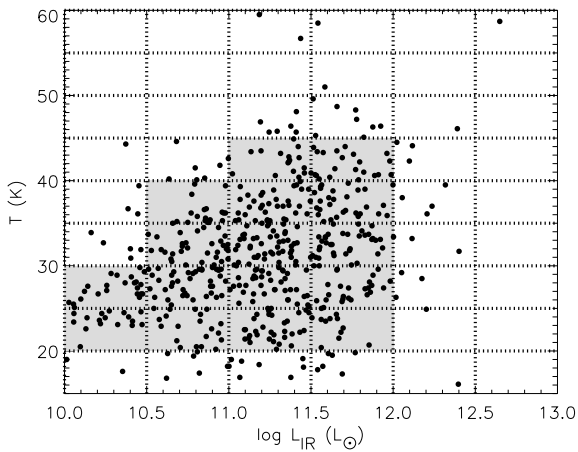, width=0.99\linewidth} 
\caption{Plot of dust temperature versus total infrared luminosity for
  the \textit{Herschel} sample. X-ray stacking is performed in the
  grey-shaded $L-T$ bins which contain 10 or more undetected sources. X-ray detected SFGs in each bin are not included in the
  stacking.}
\label{fig:LIR_temp}
\end{figure}

\subsection{Is there evolution in $L_{\rm X}/L_{\rm IR}$  with redshift?}

Fig. \ref{fig:ratio} shows the $L_{\rm SX}/L_{\rm IR}$ and $L_{\rm HX}/L_{\rm IR}$ ratio as a
function of redshift for the 3 IR luminosity classes of SFGs. 
There is no obvious change in $L_{\rm X}/L_{\rm IR}$ amongst the
\textit{Herschel} sample, especially since the
scatter in $L_{\rm X}/L_{\rm IR}$ is
likely to be similar to that of the local sample, of the order of
0.5--1\,dex. However, there seems to be an overall decrease in the
average $L_{\rm SX}/L_{\rm IR}$ for NIRGs from $z<$0.1 to $z\sim$0.8
and an increase in the average $L_{\rm HX}/L_{\rm IR}$ for ULIRGs from $z<$0.1 to
$z\sim$1.4. We believe these trends are likely due to the issues discussed in
Section \ref{sec:xray_ir_corr} rather than an intrinsic change in the
$L_{\rm X}/L_{\rm IR}$ ratio with redshift, particularly since we do
not see a trend in the LIRGs. Specifically for the NIRGs this is
reinforced by the fact that (i) the trend is not gradual
with redshift, i.e. not seen amongst the \textit{Herschel} sample and
(ii) it is only seen in the soft band. In the case of the
ULIRGs it is likely that, as previously discussed in Section
\ref{sec:xray_ir_corr}, the signal in the hard
band is boosted by AGN.

Our findings are consistent with recent results from Mineo et al. (2014\nocite{Mineo14}) who
find that the the total X-ray emission per unit SFR does not show any significant
change with redshift. On the other hand, models predict some
moderate evolution and suggest that the
$L_{\rm X}$ per unit SFR should exhibit an increase of
about a factor of 2--6 from $z=0$ to $z\sim2$ (e.g. Dijkstra et al. 2012; Fragos et
al. 2013). However, this evolution is thought to
be the result of a change in galaxy properties such as metallicity, with more
metal poor systems showing an increase in $L_{\rm X}/SFR$ (see also Linden et al. 2010\nocite{Linden10}; Kaaret et al. 2011\nocite{KSG11}). 
Although there is evidence that the mass-metallicity relation evolves
with redshift (e.g. Savaglio et al. 2005\nocite{Savaglio05}; Maiolino
et al. 2008\nocite{Maiolino08}), this is perhaps not strong enough to
translate to a measurable change in the $L_{\rm X}$ per unit SFR.

\begin{figure*}
\epsfig{file=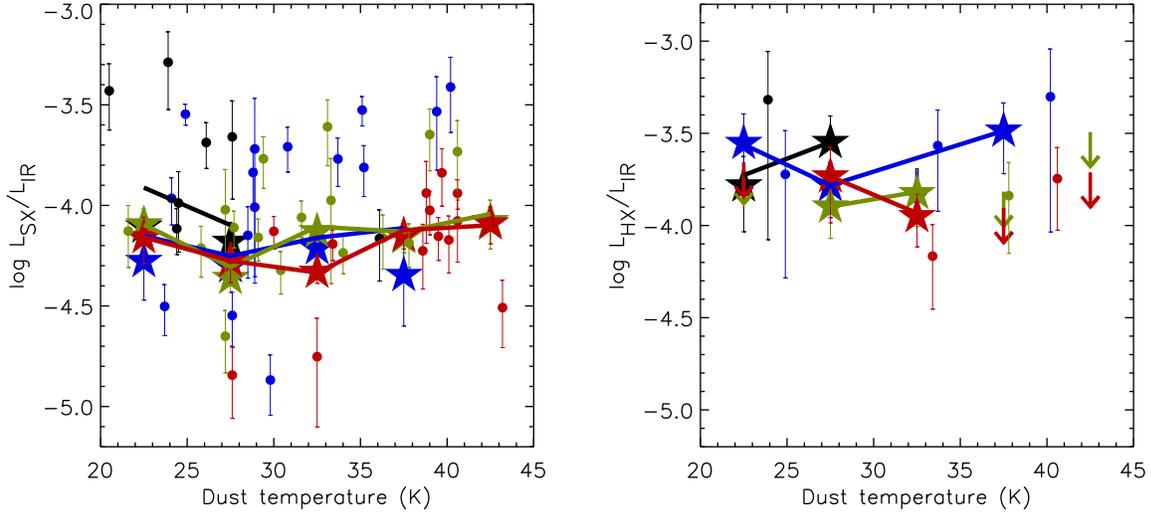, width=0.9\linewidth} 
\caption{ Plot of $L_{\rm SX}/L_{\rm IR}$ versus dust temperature for
  each grey-shaded $L-T$ bin in Fig. \ref{fig:LIR_temp}. The colour
  coding corresponds to the 4 $L_{\rm IR}$ bins: black for $10<$log$
  L_{\rm IR} <10.5$, blue for $10.5<$log $L_{\rm IR} <11$,green for
  $11<$log$L_{\rm IR} <11.5$ and red for $11.5<$log$L_{\rm IR}
  <12$. X-ray detected sources are denoted with filled circles. The
stacked values for X-ray undetected sources are denoted by large filled
stars if the stacked X-ray flux in that bin is $>3\sigma$ and upper limits if otherwise. The lines trace the average $L_{\rm SX}/L_{\rm IR}$ for all
sources per temperature bin (excluding any bins where there is an
upper limit for the stacked $L_{\rm SX}/L_{\rm IR}$ ratio). }
\label{fig:T_ratio}
\end{figure*}

\subsection{Do X-rays suffer attenuation in dusty galaxies?}
\label{sec:dust}

Our aim here is to examine whether dust obscuration 
affects the X-ray emission from star-formation. For this purpose, we
assume that for each $L_{\rm IR}$ bin (see Fig. \ref{fig:LIR_temp} for
binning), i.e. for a given SFR, the average dust temperature of galaxy is
related to physical conditions, such as gas/dust distribution. Thus lower dust temperatures are assumed to represent more extended
dust-distribution/star-forming regions,
whereas high dust temperatures are taken to represent more compact dust
distribution/star-forming regions. Within this framework, we test
the hypothesis that in galaxies with a more compact configuration, X-rays are attenuated more severely. Lehmer et
al. (2010) also investigate the effect of obscuration on the X-ray
emission using the $f_{60}/f_{100}$ colour as a proxy for dust temperature and the $L_{\rm IR}/L_{\rm
  UV}$ ratio as a proxy for extinction. They find a marginally
significant correlation and concluded that obscuration 
might be partially responsible for the deficit in the X-ray emission
at the largest SFRs in local galaxies.

Fig. \ref{fig:LIR_temp} shows dust temperature versus infrared
luminosity for the sample, split into $T - L_{\rm IR}$
bins as shown by the grid. Using the binning shown in Fig. \ref{fig:LIR_temp}, we
examine any changes in the X-ray/IR ratio as a function of dust
temperature --- see Fig. \ref{fig:T_ratio}. Interestingly, we see no correlation between the two
quantities; there is no decrease in the soft or hard X-ray
emission with increasing temperature in a given $L_{\rm IR}$ bin. This
result suggests either that (i) gas column densities in the host
galaxy are never high enough to noticeably attenuate X-ray emission or that (ii) the
regions which emit in the X-rays are not spatially coincident
with the IR-emitting regions. A situation where the latter might be the
case is if the X-rays we detect solely originate in outflows of hot gas,
in systems similar to M82 (e.g. Strickland $\&$ Stevens 2000\nocite{SS00}). However, studies of local galaxies have
demonstrated that compact sources, mainly HMXBs, which are responsible for
part of the observed soft X-ray and all of the observed hard X-ray
emission, are in fact spatially coincident with star-forming regions (e.g. Mineo et
al. 2012a). Consequently, our results more plausibly favour reason (i), i.e. that obscuration does not significantly affect the
X-rays.

\section{Summary and conclusions}
\label{sec:conclusions}
We have examined the X-ray properties of a \textit{Herschel}-selected
sample of 640 IR-luminous ($L_{\rm
  IR}>10^{10}$\,L$_{\odot}$) galaxies at $z<1.5$. As the aim of this work was to examine
X-ray emission from star-formation, AGN hosts were excluded
from our analysis (18$\pm$2 per
cent of the sample) using a set of well-established AGN criteria. 
Consequently, our final sample consisted of sources which were
considered to be star-formation dominated at all wavelengths. From
these only a small fraction (9 per cent) were detected in the X-rays. For the
remaining galaxies we calculated average luminosities through
stacking. Our aims were three-fold: (i) to revisit the X-ray/IR
correlation from star-formation, with a larger sample, extending to
lower $L_{\rm IR}$ and higher redshift using the deepest X-ray data
available, (ii) compare the X-ray properties of high redshift
star-forming galaxies to their local counterparts and (ii) identify
whether X-ray emission from star-formation is affected by the physical
properties of galaxies, such as dust obscuration. 

The soft and hard X-ray/IR correlation of star-forming galaxies was examined
over 3 orders of magnitude in  $L_{\rm IR}$,  corresponding to
2$\lesssim$SFR$\lesssim$2000\,M$_{\odot}$/yr, and at $z<1.5$. It was found to be 
approximately linear, indicating that X-ray emission is a good tracer
of star-formation, at least for sources with
sSFR$\gtrsim$0.1\,Gyr$^{-1}$. In contrast the local ($z<0.1$) relation
displays a much flatter slope, although as was discussed, it is
possible that selection biases, particularly
at low SFRs have contributed to that effect. 

We find that log\,$L_{\rm SX}/L_{\rm IR}$ for the \textit{Herschel} sample
ranges from -5.6 to -3.2, with log\,$\langle L_{\rm SX}/L_{\rm IR} \rangle=-4.3$,
whereas log\,$L_{\rm HX}/L_{\rm IR}$ extends
to higher values between -4.8 to -3, with log\,$\langle L_{\rm HX}/L_{\rm IR} \rangle=-3.8$. In both the hard and
soft bands, the log\,$L_{\rm X}/L_{\rm IR} >-3$ parameter space was found to be entirely AGN
dominated. We found that the typical $L_{\rm X}/L_{\rm IR}$
ratios of the \textit{Herschel} sources were broadly consistent with
local ($z<$0.1) SFGs, apart from discrepancies at low $L_{\rm IR}$ and
low-redshift, which we partly associate with selection biases in the local
sample. Our results showed no evidence for evolution in
the $L_{\rm X}/L_{\rm IR}$ ratio with redshift.

Finally, we addressed the question of whether X-rays are
significantly attenuated by gas/dust in the host galaxy, which would be of
particular importance in more dust-rich or compact systems. We used the
average dust temperature as a proxy, assuming that a
decrease in the average dust temperature is coupled with more extended
dust/gas distribution and star-forming regions. Interestingly, we found no evidence that
dust/gas obscuration affects X-ray emission
from star-forming regions, suggesting that gas column densities in the
host galaxy are not high enough to attenuate X-rays, unlike those
seen in AGN tori.

\section*{Acknowledgments}
This paper uses data from \textit{Herschel}'s photometers SPIRE
and PACS. SPIRE has been developed by a consortium of institutes led
by Cardiff Univ. (UK) and including: Univ. Lethbridge (Canada);
NAOC (China); CEA, LAM (France); IFSI, Univ. Padua (Italy);
IAC (Spain); Stockholm Observatory (Sweden); Imperial College
London, RAL, UCL-MSSL, UKATC, Univ. Sussex (UK); and Caltech,
JPL, NHSC, Univ. Colorado (USA). This development has been
supported by national funding agencies: CSA (Canada); NAOC
(China); CEA, CNES, CNRS (France); ASI (Italy); MCINN (Spain);
SNSB (Sweden); STFC, UKSA (UK); and NASA (USA). PACS has been
developed by a consortium of institutes led by MPE (Germany) and including UVIE (Austria); KU Leuven, CSL, IMEC (Belgium);
CEA, LAM (France); MPIA (Germany); INAF-IFSI/OAA/OAP/OAT, LENS,
SISSA (Italy); IAC (Spain). This development has been supported by the funding
agencies BMVIT (Austria), ESA-PRODEX (Belgium), CEA/CNES (France),
DLR (Germany), ASI/INAF (Italy), and CICYT/MCYT (Spain). The
scientific results reported in this article are based to a significant
degree on observations made by the Chandra X-ray Observatory.

\bibliographystyle{mn2e}
\bibliography{references}

\begin{thebibliography}{}

\bibitem[\protect\citeauthoryear{{Akiyama}, {Ueda}, {Ohta}, {Takahashi} \&
  {Yamada}}{{Akiyama} et~al.}{2003}]{Akiyama03}
{Akiyama} M.,  {Ueda} Y.,  {Ohta} K.,  {Takahashi} T.,    {Yamada} T.,  2003,
  \apjs, 148, 275

\bibitem[\protect\citeauthoryear{{Alexander} et~al.,}{{Alexander}
  et~al.}{2003}]{Alexander03b}
{Alexander} D.~M.,  et~al., 2003, \aj, 126, 539

\bibitem[\protect\citeauthoryear{{Bendo} et~al.,}{{Bendo}
  et~al.}{2010}]{Bendo10}
{Bendo} G.~J.,  et~al., 2010, \aap, 518, L65

\bibitem[\protect\citeauthoryear{{Berta} et~al.,}{{Berta}
  et~al.}{2011}]{Berta11}
{Berta} S.,  et~al., 2011, \aap, 532, A49

\bibitem[\protect\citeauthoryear{{Brandt} \& {Hasinger}}{{Brandt} \&
  {Hasinger}}{2005}]{BH05}
{Brandt} W.~N.,  {Hasinger} G.,  2005, \araa, 43, 827

\bibitem[\protect\citeauthoryear{{Bruzual} \& {Charlot}}{{Bruzual} \&
  {Charlot}}{2003}]{BC03}
{Bruzual} G.,  {Charlot} S.,  2003, \mnras, 344, 1000

\bibitem[\protect\citeauthoryear{{Calzetti}, {Armus}, {Bohlin}, {Kinney},
  {Koornneef} \& {Storchi-Bergmann}}{{Calzetti} et~al.}{2000}]{Calzetti00}
{Calzetti} D.,  {Armus} L.,  {Bohlin} R.~C.,  {Kinney} A.~L.,  {Koornneef} J.,
    {Storchi-Bergmann} T.,  2000, \apj, 533, 682

\bibitem[\protect\citeauthoryear{{Cardamone} et~al.,}{{Cardamone}
  et~al.}{2010}]{Cardamone10}
{Cardamone} C.~N.,  et~al., 2010, \apjs, 189, 270

\bibitem[\protect\citeauthoryear{{Colbert}, {Heckman}, {Ptak}, {Strickland} \&
  {Weaver}}{{Colbert} et~al.}{2004}]{Colbert04}
{Colbert} E.~J.~M.,  {Heckman} T.~M.,  {Ptak} A.~F.,  {Strickland} D.~K.,
  {Weaver} K.~A.,  2004, \apj, 602, 231

\bibitem[\protect\citeauthoryear{{Comastri} et~al.,}{{Comastri}
  et~al.}{2011}]{Comastri11}
{Comastri} A.,  et~al., 2011, \aap, 526, L9

\bibitem[\protect\citeauthoryear{{Comastri}, {Setti}, {Zamorani} \&
  {Hasinger}}{{Comastri} et~al.}{1995}]{Comastri95}
{Comastri} A.,  {Setti} G.,  {Zamorani} G.,    {Hasinger} G.,  1995, \aap, 296,
  1

\bibitem[\protect\citeauthoryear{{Coppin} et~al.,}{{Coppin}
  et~al.}{2008}]{Coppin08}
{Coppin} K.,  et~al., 2008, \mnras, 384, 1597

\bibitem[\protect\citeauthoryear{{David}, {Jones} \& {Forman}}{{David}
  et~al.}{1992}]{DJF92}
{David} L.~P.,  {Jones} C.,    {Forman} W.,  1992, \apj, 388, 82

\bibitem[\protect\citeauthoryear{{Donley} et~al.,}{{Donley}
  et~al.}{2012}]{Donley12}
{Donley} J.~L.,  et~al., 2012, \apj, 748, 142

\bibitem[\protect\citeauthoryear{{Fabbiano}}{{Fabbiano}}{1988}]{Fabbiano88}
{Fabbiano} G.,  1988, \apj, 325, 544

\bibitem[\protect\citeauthoryear{{Fabbiano}}{{Fabbiano}}{1989}]{Fabbiano89}
{Fabbiano} G.,  1989, \araa, 27, 87

\bibitem[\protect\citeauthoryear{{Fabbiano}}{{Fabbiano}}{2005}]{Fabbiano05}
{Fabbiano} G.,  2005, Science, 307, 533

\bibitem[\protect\citeauthoryear{{Fabbiano} et~al.,}{{Fabbiano}
  et~al.}{2004}]{Fabbiano04}
{Fabbiano} G.,  et~al., 2004, \apjl, 605, L21

\bibitem[\protect\citeauthoryear{{Fabbiano}, {Schweizer} \&
  {Mackie}}{{Fabbiano} et~al.}{1997}]{FSM97}
{Fabbiano} G.,  {Schweizer} F.,    {Mackie} G.,  1997, \apj, 478, 542

\bibitem[\protect\citeauthoryear{{Fabbiano} \& {Trinchieri}}{{Fabbiano} \&
  {Trinchieri}}{1984}]{FT84}
{Fabbiano} G.,  {Trinchieri} G.,  1984, \apj, 286, 491

\bibitem[\protect\citeauthoryear{{Farrah} et~al.,}{{Farrah}
  et~al.}{2008}]{Farrah08}
{Farrah} D.,  et~al., 2008, \apj, 677, 957

\bibitem[\protect\citeauthoryear{{Franceschini} et~al.,}{{Franceschini}
  et~al.}{2003}]{Franceschini03b}
{Franceschini} A.,  et~al., 2003, \mnras, 343, 1181

\bibitem[\protect\citeauthoryear{{Georgakakis} et~al.,}{{Georgakakis}
  et~al.}{2008}]{Georgakakis08a}
{Georgakakis} A.,  et~al., 2008, \mnras, 385, 2049

\bibitem[\protect\citeauthoryear{{Georgakakis}, {Rowan-Robinson}, {Babbedge} \&
  {Georgantopoulos}}{{Georgakakis} et~al.}{2007}]{Georgakakis07}
{Georgakakis} A.,  {Rowan-Robinson} M.,  {Babbedge} T.~S.~R.,
  {Georgantopoulos} I.,  2007, \mnras, 377, 203

\bibitem[\protect\citeauthoryear{{Georgantopoulos}, {Georgakakis} \&
  {Koulouridis}}{{Georgantopoulos} et~al.}{2005}]{GGK05}
{Georgantopoulos} I.,  {Georgakakis} A.,    {Koulouridis} E.,  2005, \mnras,
  360, 782

\bibitem[\protect\citeauthoryear{{Ghosh} \& {White}}{{Ghosh} \&
  {White}}{2001}]{GW01}
{Ghosh} P.,  {White} N.~E.,  2001, \apjl, 559, L97

\bibitem[\protect\citeauthoryear{{Giavalisco} et~al.,}{{Giavalisco}
  et~al.}{2004}]{Giavalisco04}
{Giavalisco} M.,  et~al., 2004, \apjl, 600, L93

\bibitem[\protect\citeauthoryear{{Gilfanov}, {Grimm} \& {Sunyaev}}{{Gilfanov}
  et~al.}{2004}]{GGS04}
{Gilfanov} M.,  {Grimm} H.,    {Sunyaev} R.,  2004, \mnras, 347, L57

\bibitem[\protect\citeauthoryear{{Griffin} et~al.,}{{Griffin}
  et~al.}{2010}]{Griffin10}
{Griffin} M.~J.,  et~al., 2010, \aap, 518, L3+

\bibitem[\protect\citeauthoryear{{Griffiths} \& {Padovani}}{{Griffiths} \&
  {Padovani}}{1990}]{GP90}
{Griffiths} R.~E.,  {Padovani} P.,  1990, \apj, 360, 483

\bibitem[\protect\citeauthoryear{{Griffiths}, {Ptak}, {Feigelson}, {Garmire},
  {Townsley}, {Brandt}, {Sambruna} \& {Bregman}}{{Griffiths}
  et~al.}{2000}]{Griffiths00}
{Griffiths} R.~E.,  {Ptak} A.,  {Feigelson} E.~D.,  {Garmire} G.,  {Townsley}
  L.,  {Brandt} W.~N.,  {Sambruna} R.,    {Bregman} J.~N.,  2000, Science, 290,
  1325

\bibitem[\protect\citeauthoryear{{Grimes}, {Heckman}, {Strickland} \&
  {Ptak}}{{Grimes} et~al.}{2005}]{Grimes05}
{Grimes} J.~P.,  {Heckman} T.,  {Strickland} D.,    {Ptak} A.,  2005, \apj,
  628, 187

\bibitem[\protect\citeauthoryear{{Grimm}, {Gilfanov} \& {Sunyaev}}{{Grimm}
  et~al.}{2002}]{GGS02}
{Grimm} H.,  {Gilfanov} M.,    {Sunyaev} R.,  2002, \aap, 391, 923

\bibitem[\protect\citeauthoryear{{Grimm}, {Gilfanov} \& {Sunyaev}}{{Grimm}
  et~al.}{2003}]{GGS03}
{Grimm} H.,  {Gilfanov} M.,    {Sunyaev} R.,  2003, \mnras, 339, 793

\bibitem[\protect\citeauthoryear{{Hasinger}, {Miyaji} \& {Schmidt}}{{Hasinger}
  et~al.}{2005}]{Hasinger05}
{Hasinger} G.,  {Miyaji} T.,    {Schmidt} M.,  2005, \aap, 441, 417

\bibitem[\protect\citeauthoryear{{Hornschemeier}, {Bauer}, {Alexander},
  {Brandt}, {Sargent}, {Bautz}, {Conselice}, {Garmire}, {Schneider} \&
  {Wilson}}{{Hornschemeier} et~al.}{2003}]{Hornschemeier03}
{Hornschemeier} A.~E.,  {Bauer} F.~E.,  {Alexander} D.~M.,  {Brandt} W.~N.,
  {Sargent} W.~L.~W.,  {Bautz} M.~W.,  {Conselice} C.,  {Garmire} G.~P.,
  {Schneider} D.~P.,    {Wilson} G.,  2003, \aj, 126, 575

\bibitem[\protect\citeauthoryear{{Hornschemeier}, {Brandt}, {Alexander},
  {Bauer}, {Garmire}, {Schneider}, {Bautz} \& {Chartas}}{{Hornschemeier}
  et~al.}{2002}]{Hornschemeier02}
{Hornschemeier} A.~E.,  {Brandt} W.~N.,  {Alexander} D.~M.,  {Bauer} F.~E.,
  {Garmire} G.~P.,  {Schneider} D.~P.,  {Bautz} M.~W.,    {Chartas} G.,  2002,
  \apj, 568, 82

\bibitem[\protect\citeauthoryear{{Kaaret}, {Schmitt} \& {Gorski}}{{Kaaret}
  et~al.}{2011}]{KSG11}
{Kaaret} P.,  {Schmitt} J.,    {Gorski} M.,  2011, \apj, 741, 10

\bibitem[\protect\citeauthoryear{{Kennicutt}
  Jr.}{{Kennicutt}}{1998}]{Kennicutt98}
{Kennicutt} Jr. R.~C.,  1998, \araa, 36, 189

\bibitem[\protect\citeauthoryear{{Kriek}, {van Dokkum}, {Labb{\'e}}, {Franx},
  {Illingworth}, {Marchesini} \& {Quadri}}{{Kriek} et~al.}{2009}]{Kriek09}
{Kriek} M.,  {van Dokkum} P.~G.,  {Labb{\'e}} I.,  {Franx} M.,  {Illingworth}
  G.~D.,  {Marchesini} D.,    {Quadri} R.~F.,  2009, \apj, 700, 221

\bibitem[\protect\citeauthoryear{{Laird} et~al.,}{{Laird}
  et~al.}{2009}]{Laird09}
{Laird} E.~S.,  et~al., 2009, \apjs, 180, 102

\bibitem[\protect\citeauthoryear{{Lehmer}, {Alexander}, {Bauer}, {Brandt},
  {Goulding}, {Jenkins}, {Ptak} \& {Roberts}}{{Lehmer} et~al.}{2010}]{Lehmer10}
{Lehmer} B.~D.,  {Alexander} D.~M.,  {Bauer} F.~E.,  {Brandt} W.~N.,
  {Goulding} A.~D.,  {Jenkins} L.~P.,  {Ptak} A.,    {Roberts} T.~P.,  2010,
  ArXiv e-prints

\bibitem[\protect\citeauthoryear{{Linden}, {Kalogera}, {Sepinsky}, {Prestwich},
  {Zezas} \& {Gallagher}}{{Linden} et~al.}{2010}]{Linden10}
{Linden} T.,  {Kalogera} V.,  {Sepinsky} J.~F.,  {Prestwich} A.,  {Zezas} A.,
   {Gallagher} J.~S.,  2010, \apj, 725, 1984

\bibitem[\protect\citeauthoryear{{Lo Faro} et~al.,}{{Lo Faro}
  et~al.}{2013}]{LoFaro13}
{Lo Faro} B.,  et~al., 2013, \apj, 762, 108

\bibitem[\protect\citeauthoryear{{Lutz} et~al.,}{{Lutz}  et~al.}{2011}]{Lutz11}
{Lutz} D.,  et~al., 2011, \aap, 532, A90

\bibitem[\protect\citeauthoryear{{Magnelli}, {Elbaz}, {Chary}, {Dickinson}, {Le
  Borgne}, {Frayer} \& {Willmer}}{{Magnelli} et~al.}{2009}]{Magnelli09}
{Magnelli} B.,  {Elbaz} D.,  {Chary} R.~R.,  {Dickinson} M.,  {Le Borgne} D.,
  {Frayer} D.~T.,    {Willmer} C.~N.~A.,  2009, \aap, 496, 57

\bibitem[\protect\citeauthoryear{{Maiolino} et~al.,}{{Maiolino}
  et~al.}{2008}]{Maiolino08}
{Maiolino} R.,  et~al., 2008, \aap, 488, 463

\bibitem[\protect\citeauthoryear{{Mas-Hesse}, {Ot{\'{\i}}-Floranes} \&
  {Cervi{\~n}o}}{{Mas-Hesse} et~al.}{2008}]{MOC08}
{Mas-Hesse} J.~M.,  {Ot{\'{\i}}-Floranes} H.,    {Cervi{\~n}o} M.,  2008, \aap,
  483, 71

\bibitem[\protect\citeauthoryear{{Mineo}, {Gilfanov}, {Lehmer}, {Morrison} \&
  {Sunyaev}}{{Mineo} et~al.}{2014}]{Mineo14}
{Mineo} S.,  {Gilfanov} M.,  {Lehmer} B.~D.,  {Morrison} G.~E.,    {Sunyaev}
  R.,  2014, \mnras, 437, 1698

\bibitem[\protect\citeauthoryear{{Mineo}, {Gilfanov} \& {Sunyaev}}{{Mineo}
  et~al.}{2012a}]{MGS12a}
{Mineo} S.,  {Gilfanov} M.,    {Sunyaev} R.,  2012a, \mnras, 419, 2095

\bibitem[\protect\citeauthoryear{{Mineo}, {Gilfanov} \& {Sunyaev}}{{Mineo}
  et~al.}{2012b}]{MGS12b}
{Mineo} S.,  {Gilfanov} M.,    {Sunyaev} R.,  2012b, \mnras, 426, 1870

\bibitem[\protect\citeauthoryear{{Nguyen}}{{Nguyen}}{2010}]{Nguyen10}
{Nguyen} H.~T.~o.,  2010, \aap, 518, L5+

\bibitem[\protect\citeauthoryear{{Oliver} et~al.,}{{Oliver}
  et~al.}{2012}]{Oliver12}
{Oliver} S.~J.,  et~al., 2012, \mnras, 424, 1614

\bibitem[\protect\citeauthoryear{{Page}, {Mason}, {McHardy}, {Jones} \&
  {Carrera}}{{Page} et~al.}{1997}]{Page97}
{Page} M.~J.,  {Mason} K.~O.,  {McHardy} I.~M.,  {Jones} L.~R.,    {Carrera}
  F.~J.,  1997, \mnras, 291, 324

\bibitem[\protect\citeauthoryear{{Paolillo}, {Schreier}, {Giacconi},
  {Koekemoer} \& {Grogin}}{{Paolillo} et~al.}{2004}]{Paolillo04}
{Paolillo} M.,  {Schreier} E.~J.,  {Giacconi} R.,  {Koekemoer} A.~M.,
  {Grogin} N.~A.,  2004, \apj, 611, 93

\bibitem[\protect\citeauthoryear{{Persic}, {Rephaeli}, {Braito}, {Cappi},
  {Della Ceca}, {Franceschini} \& {Gruber}}{{Persic} et~al.}{2004}]{Persic04}
{Persic} M.,  {Rephaeli} Y.,  {Braito} V.,  {Cappi} M.,  {Della Ceca} R.,
  {Franceschini} A.,    {Gruber} D.~E.,  2004, \aap, 419, 849

\bibitem[\protect\citeauthoryear{{Pilbratt} et~al.,}{{Pilbratt}
  et~al.}{2010}]{Pilbratt10}
{Pilbratt} G.~L.,  et~al., 2010, \aap, 518, L1+

\bibitem[\protect\citeauthoryear{{Poglitsch} et~al.,}{{Poglitsch}
  et~al.}{2010}]{Poglitsch10}
{Poglitsch} A.,  et~al., 2010, \aap, 518, L2+

\bibitem[\protect\citeauthoryear{{Ptak}, {Griffiths}, {White} \&
  {Ghosh}}{{Ptak} et~al.}{2001}]{Ptak01}
{Ptak} A.,  {Griffiths} R.,  {White} N.,    {Ghosh} P.,  2001, \apjl, 559, L91

\bibitem[\protect\citeauthoryear{{Ranalli}, {Comastri} \& {Setti}}{{Ranalli}
  et~al.}{2003}]{RCS03}
{Ranalli} P.,  {Comastri} A.,    {Setti} G.,  2003, \aap, 399, 39

\bibitem[\protect\citeauthoryear{{Ranalli}, {Comastri} \& {Setti}}{{Ranalli}
  et~al.}{2005}]{RCS05}
{Ranalli} P.,  {Comastri} A.,    {Setti} G.,  2005, \aap, 440, 23

\bibitem[\protect\citeauthoryear{{Rosa-Gonz{\'a}lez}, {Burgarella}, {Nandra},
  {Kunth}, {Terlevich} \& {Terlevich}}{{Rosa-Gonz{\'a}lez}
  et~al.}{2007}]{Rosa-Gonzalez07}
{Rosa-Gonz{\'a}lez} D.,  {Burgarella} D.,  {Nandra} K.,  {Kunth} D.,
  {Terlevich} E.,    {Terlevich} R.,  2007, \mnras, 379, 357

\bibitem[\protect\citeauthoryear{{Roseboom} et~al.,}{{Roseboom}
  et~al.}{2010}]{Roseboom10}
{Roseboom} I.~G.,  et~al., 2010, \mnras, pp 1630--+

\bibitem[\protect\citeauthoryear{{Roseboom} et~al.,}{{Roseboom}
  et~al.}{2012}]{Roseboom12}
{Roseboom} I.~G.,  et~al., 2012, \mnras, 419, 2758

\bibitem[\protect\citeauthoryear{{Sajina} et~al.,}{{Sajina}
  et~al.}{2008}]{Sajina08}
{Sajina} A.,  et~al., 2008, \apj, 683, 659

\bibitem[\protect\citeauthoryear{{Sanders} \& {Mirabel}}{{Sanders} \&
  {Mirabel}}{1996}]{SM96}
{Sanders} D.~B.,  {Mirabel} I.~F.,  1996, \araa, 34, 749

\bibitem[\protect\citeauthoryear{{Santini} et~al.,}{{Santini}
  et~al.}{2009}]{Santini09}
{Santini} P.,  et~al., 2009, \aap, 504, 751

\bibitem[\protect\citeauthoryear{{Savaglio} et~al.,}{{Savaglio}
  et~al.}{2005}]{Savaglio05}
{Savaglio} S.,  et~al., 2005, \apj, 635, 260

\bibitem[\protect\citeauthoryear{{Seymour} et~al.,}{{Seymour}
  et~al.}{2011}]{Seymour11}
{Seymour} N.,  et~al., 2011, \mnras, 413, 1777

\bibitem[\protect\citeauthoryear{{Soifer} et~al.,}{{Soifer}
  et~al.}{1984}]{Soifer84a}
{Soifer} B.~T.,  et~al., 1984, \apjl, 278, L71

\bibitem[\protect\citeauthoryear{{Soifer}, {Neugebauer} \& {Houck}}{{Soifer}
  et~al.}{1987}]{SNH87}
{Soifer} B.~T.,  {Neugebauer} G.,    {Houck} J.~R.,  1987, \araa, 25, 187

\bibitem[\protect\citeauthoryear{{Soria}, {Hau}, {Graham}, {Kong}, {Kuin},
  {Li}, {Liu} \& {Wu}}{{Soria} et~al.}{2010}]{Soria10}
{Soria} R.,  {Hau} G.~K.~T.,  {Graham} A.~W.,  {Kong} A.~K.~H.,  {Kuin}
  N.~P.~M.,  {Li} I.-H.,  {Liu} J.-F.,    {Wu} K.,  2010, \mnras, 405, 870

\bibitem[\protect\citeauthoryear{{Soria}, {Kuntz}, {Winkler}, {Blair}, {Long},
  {Plucinsky} \& {Whitmore}}{{Soria} et~al.}{2012}]{Soria12}
{Soria} R.,  {Kuntz} K.~D.,  {Winkler} P.~F.,  {Blair} W.~P.,  {Long} K.~S.,
  {Plucinsky} P.~P.,    {Whitmore} B.~C.,  2012, \apj, 750, 152

\bibitem[\protect\citeauthoryear{{Strickland}, {Heckman}, {Colbert}, {Hoopes}
  \& {Weaver}}{{Strickland} et~al.}{2004}]{Strickland04}
{Strickland} D.~K.,  {Heckman} T.~M.,  {Colbert} E.~J.~M.,  {Hoopes} C.~G.,
  {Weaver} K.~A.,  2004, \apjs, 151, 193

\bibitem[\protect\citeauthoryear{{Strickland} \& {Stevens}}{{Strickland} \&
  {Stevens}}{2000}]{SS00}
{Strickland} D.~K.,  {Stevens} I.~R.,  2000, \mnras, 314, 511

\bibitem[\protect\citeauthoryear{{Symeonidis} et~al.,}{{Symeonidis}
  et~al.}{2011}]{Symeonidis11b}
{Symeonidis} M.,  et~al., 2011, \mnras, 417, 2239

\bibitem[\protect\citeauthoryear{{Symeonidis} et~al.,}{{Symeonidis}
  et~al.}{2013}]{Symeonidis13a}
{Symeonidis} M.,  et~al., 2013, \mnras, 431, 2317

\bibitem[\protect\citeauthoryear{{Symeonidis}, {Page}, {Seymour}, {Dwelly},
  {Coppin}, {McHardy}, {Rieke} \& {Huynh}}{{Symeonidis}
  et~al.}{2009}]{Symeonidis09}
{Symeonidis} M.,  {Page} M.~J.,  {Seymour} N.,  {Dwelly} T.,  {Coppin} K.,
  {McHardy} I.,  {Rieke} G.~H.,    {Huynh} M.,  2009, \mnras, 397, 1728

\bibitem[\protect\citeauthoryear{{Tozzi} et~al.,}{{Tozzi}
  et~al.}{2006}]{Tozzi06}
{Tozzi} P.,  et~al., 2006, \aap, 451, 457

\bibitem[\protect\citeauthoryear{{Tueller} et~al.,}{{Tueller}
  et~al.}{2010}]{Tueller10}
{Tueller} J.,  et~al., 2010, \apjs, 186, 378

\bibitem[\protect\citeauthoryear{{U} et~al.,}{{U}  et~al.}{2012}]{U12}
{U} V.,  et~al., 2012, \apjs, 203, 9

\bibitem[\protect\citeauthoryear{{Vattakunnel} et~al.,}{{Vattakunnel}
  et~al.}{2012}]{Vattakunnel12}
{Vattakunnel} S.,  et~al., 2012, \mnras, 420, 2190

\bibitem[\protect\citeauthoryear{{Villforth}, {Koekemoer} \&
  {Grogin}}{{Villforth} et~al.}{2010}]{Villforth10}
{Villforth} C.,  {Koekemoer} A.~M.,    {Grogin} N.~A.,  2010, \apj, 723, 737

\bibitem[\protect\citeauthoryear{{Villforth}, {Sarajedini} \&
  {Koekemoer}}{{Villforth} et~al.}{2012}]{Villforth12}
{Villforth} C.,  {Sarajedini} V.,    {Koekemoer} A.,  2012, \mnras, 426, 360

\bibitem[\protect\citeauthoryear{{Xue} et~al.,}{{Xue}  et~al.}{2011}]{Xue11}
{Xue} Y.~Q.,  et~al., 2011, \apjs, 195, 10

\bibitem[\protect\citeauthoryear{{Young} et~al.,}{{Young}
  et~al.}{2012}]{Young12}
{Young} M.,  et~al., 2012, \apj, 748, 124

\end{thebibliography}

\clearpage
\appendix

\section{AGN impact}
\label{appendixA}

As our AGN selection is not complete, in the sense that we cannot find
all AGN in the sample, there is a possibility
that some of our `SFG'-classified sources host an AGN. Here we
investigate whether this would have an impact on the derivation of the X-ray/IR correlations investigated in Section
\ref{sec:xray_ir_corr}, see also Fig. \ref{fig:corr}. 

\begin{figure*}
\begin{tabular}{c|c}
\epsfig{file=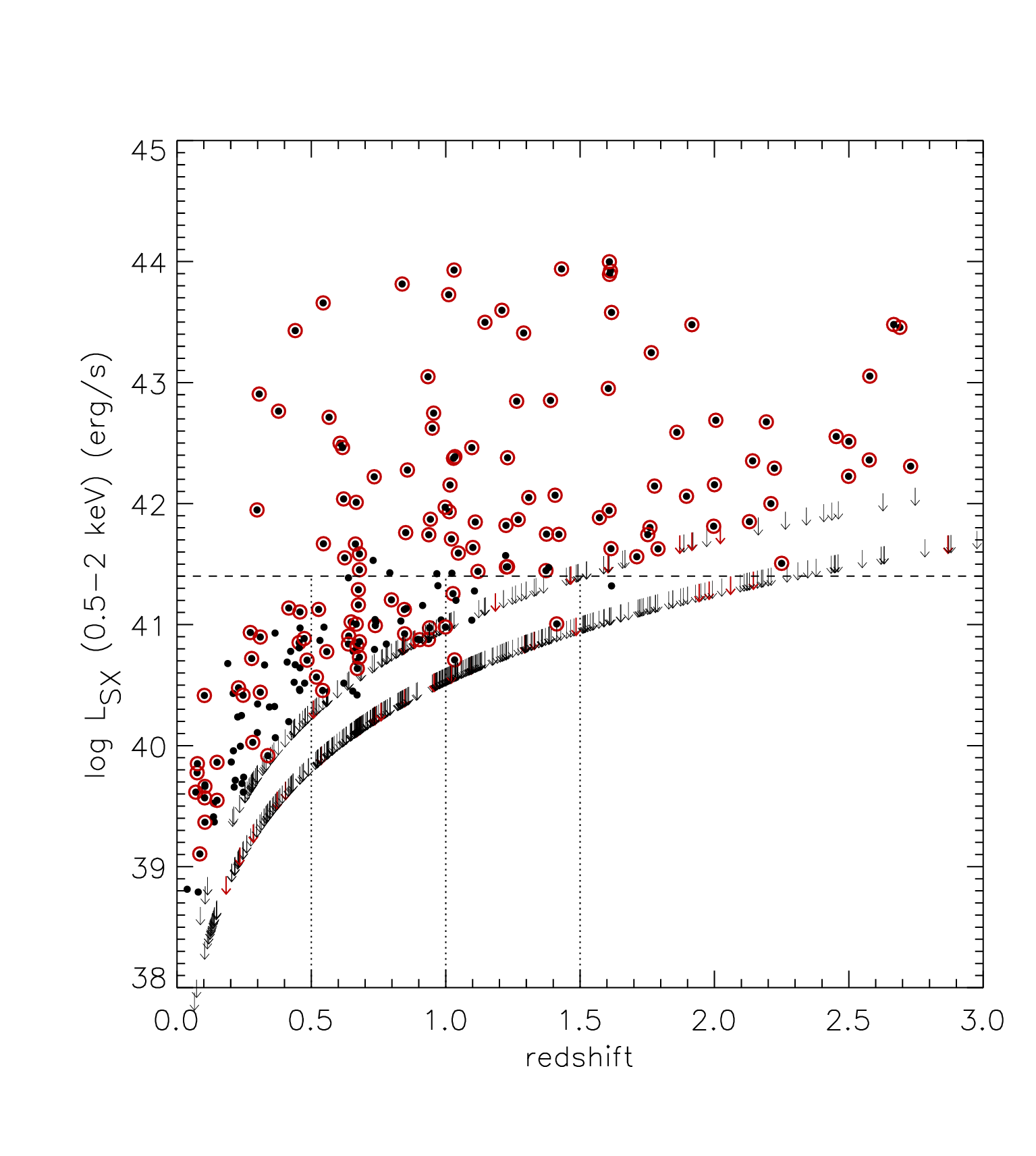, width=0.4\linewidth} & \epsfig{file=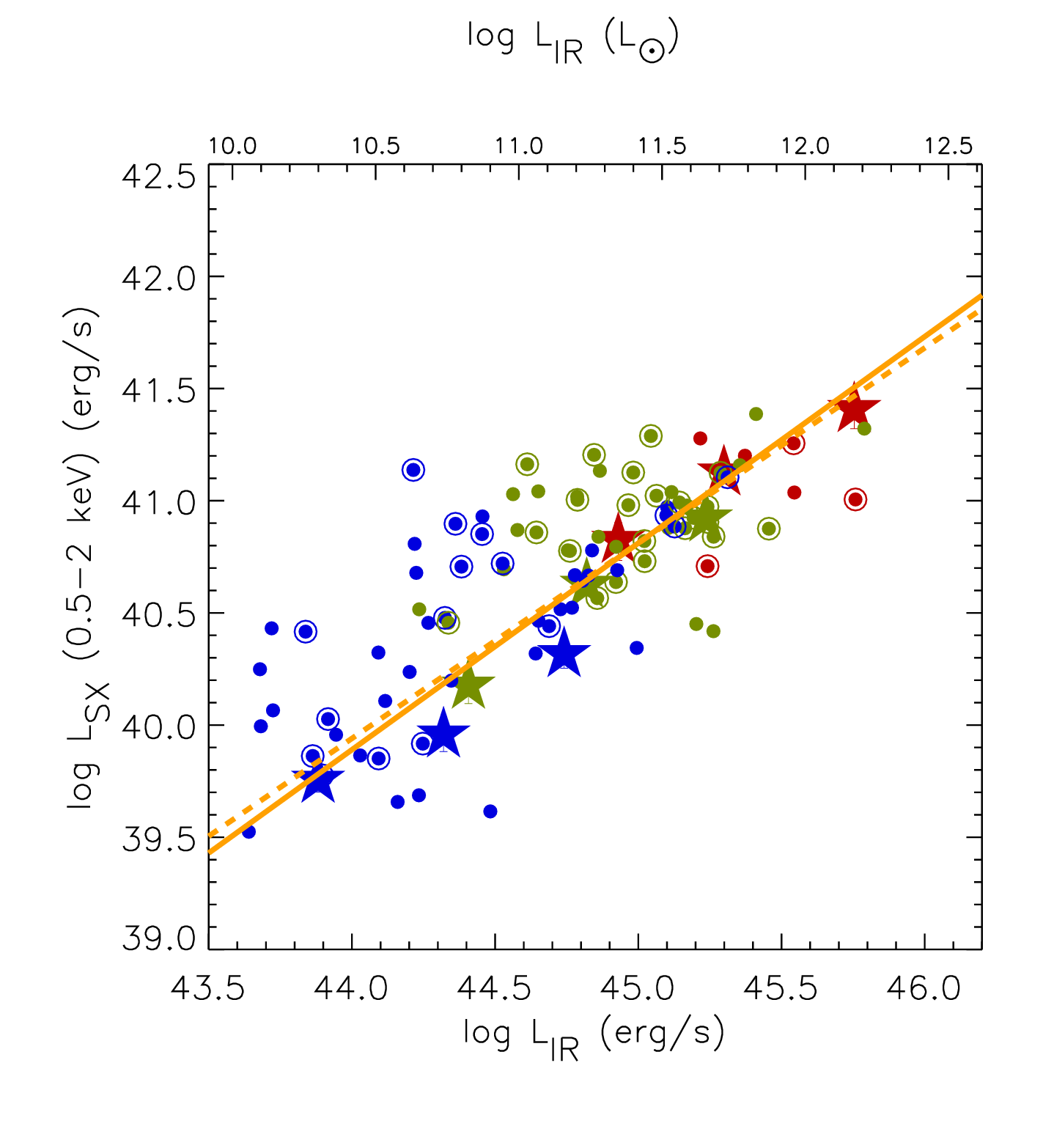, width=0.4\linewidth} \\
\end{tabular}
\caption{\textit{Left panel}: soft X-ray luminosity ($L_{\rm SX}$) versus
  redshift. The vertical dotted lines indicate the 3 redshift
  bins. The horizontal dotted line is at log\,$L_{SX}$=41.4 roughly
  corresponding to the sensitivity threshold
  of the shallowest X-ray survey at $z\sim$1.5. \textit{Right panel}: soft X-ray luminosity ($L_{\rm SX}$) versus
  $L_{\rm IR}$, a remake of Fig. \ref{fig:corr} upper panel including
  AGN. SFGs are denoted with filled circles and AGN with log\,$L_{SX} <
41.4$ as filled circles with circular outline. The colour coding is: blue for
$z<0.5$, green for 0.5$<z<$1 and red for 1$<z<$1.5. Large filled stars
correspond to the stacked values if they are $>3\sigma$ (otherwise
shown as upper limits), whereas the points correspond to the detections. 
A linear fit to the arithmetic weighted mean (in 5 $L_{\rm IR}$ bins) of the stacking and
  detections is shown as a solid orange line for the \textit{Herschel}
  SFGs only and a dotted orange line for SFGs plus AGN with log\,$L_{SX} <
41.4$. } 
\label{fig:appendix_fig1}
\end{figure*}

\begin{figure*}
\begin{tabular}{c|c}
\epsfig{file=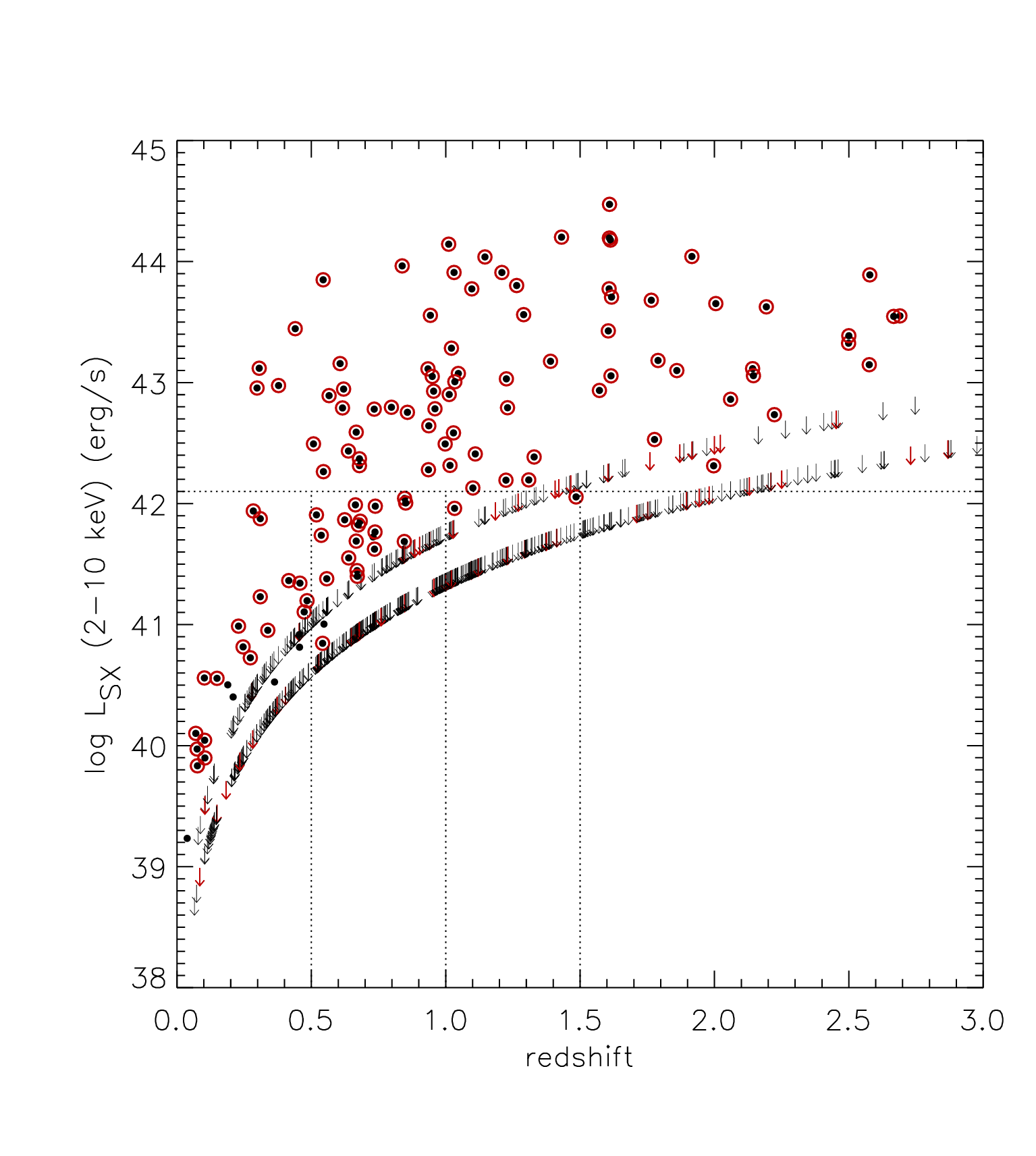, width=0.4\linewidth} & \epsfig{file=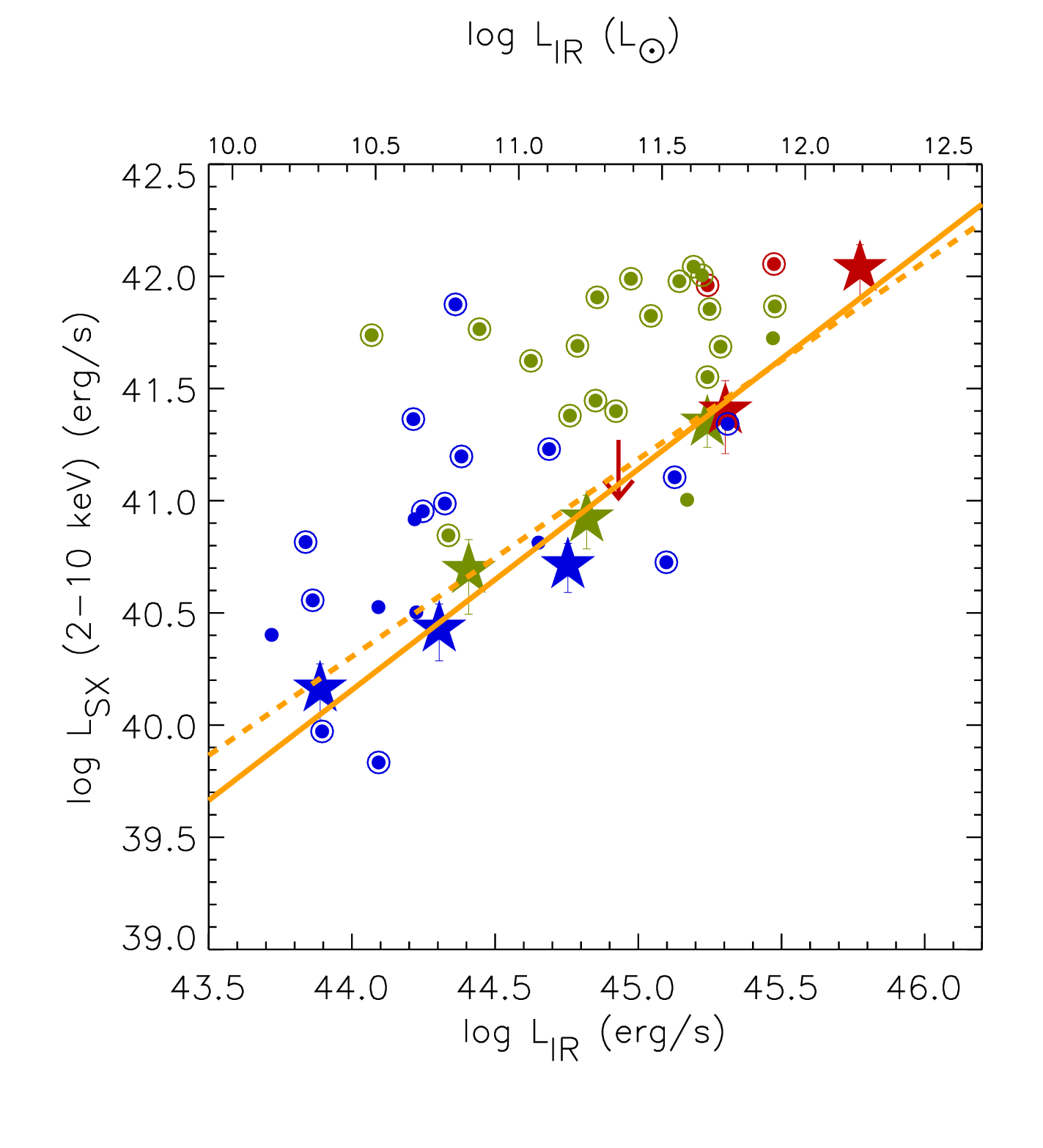, width=0.4\linewidth} \\
\end{tabular}
\caption{\textit{Left panel}: hard X-ray luminosity ($L_{\rm HX}$) versus
  redshift. The vertical dotted lines indicate the 3 redshift
  bins. The horizontal dotted line is at log\,$L_{HX}$=42.1 roughly
  corresponding to the sensitivity threshold
  of the shallowest X-ray survey at $z\sim$1.5. \textit{Right panel}: hard X-ray luminosity ($L_{\rm HX}$) versus
  $L_{\rm IR}$, a remake of Fig. \ref{fig:corr} lower panel including
  AGN. SFGs are denoted with filled circles and AGN with log\,$L_{SX} <
42.1$ as filled circles with circular outline. The colour coding is: blue for
$z<0.5$, green for 0.5$<z<$1 and red for 1$<z<$1.5. Large filled stars
correspond to the stacked values if they are $>3\sigma$ (otherwise
shown as upper limits), whereas the points correspond to the detections. A linear fit to the arithmetic weighted mean (in 5 $L_{\rm IR}$ bins) of the stacking and
  detections is shown as a solid orange line for the \textit{Herschel}
  SFGs only and a dotted orange line for SFGs plus AGN with log\,$L_{SX} <
42.1$. } 
\label{fig:appendix_fig2}
\end{figure*}

The left panels of Figs \ref{fig:appendix_fig1} and
\ref{fig:appendix_fig2}, show the soft and hard X-ray luminosity of
the sample as a function of redshift. Note that because of the
surveys' detection limits, X-ray detected sources in the high redshift
bin are invariably more luminous than those in the low redshift
bin. In addition, low luminosity AGN identified in the low redshift
bin, are below the detection threshold at high redshift and it is
likely that they would creep into our sample of X-ray undetected SFGs
and contaminate the average X-ray luminosity of the high redshift bin. 
To investigate this effect we perform the following simple experiment: for the soft X-rays, we re-examine the
X-ray/IR correlation including all AGN with log\,$L_{\rm SX} <
41.4$, which is the detection threshold of the highest redshift bin used in
our analysis ($z<1.5$), as shown in the left panel of Fig. \ref{fig:appendix_fig1}. Similarly, we re-examine the hard X-ray/IR
correlation by including all AGN with log\,$L_{\rm HX} <42.1$ (see left panel of
\ref{fig:appendix_fig2}). 
The right panels of Figs \ref{fig:appendix_fig1} and
\ref{fig:appendix_fig2} demonstrate how the X-ray/IR correlation, shown in
Fig. \ref{fig:corr}, would change. For the soft band (right panel of
Fig. \ref{fig:appendix_fig1}), there
is only a marginal increase in the X-ray luminosity for the low redshift
bin, not surprising, as the AGN now included have
comparable luminosity to X-ray detected sources classified as SFGs, so
they are not expected to increase the average X-ray luminosity. 
For the hard band (right panel of
Fig. \ref{fig:appendix_fig2}), there is a more significant increase in
the average X-ray luminosity for the low redshift
bins; the average $L_{\rm HX}$ per $L_{\rm IR}$ increases
by up to 0.26\,dex. The larger increase in $L_{\rm HX}$ per $L_{\rm IR}$
is to be expected as the hard band is lacking in X-ray detected
SFGs. 
In line with these results, we can now assume that for the high redshift bin, the soft band
luminosity is only slightly boosted, and the hard-band luminosity is
potentially boosted by
up to 0.26\,dex. This means that the hard X-ray/IR correlation slope
is potentially flatter than what we measured in section \ref{sec:xray_ir_corr}, with the slope changing from 0.98 to 0.75.

\label{lastpage}

\end{document}